\documentclass[letterpaper,11pt]{article}
\usepackage{amssymb}
\usepackage{amsthm}
\usepackage{amsmath}
\usepackage{mathrsfs,txfonts,paralist,verbatim,fullpage}
\usepackage{hyperref}
\usepackage{graphicx}

\DeclareSymbolFont{symbolsC}{U}{txsyc}{m}{n}
\SetSymbolFont{symbolsC}{bold}{U}{txsyc}{bx}{n}
\DeclareFontSubstitution{U}{txsyc}{m}{n}
\DeclareMathSymbol{\coloneqq}{\mathrel}{symbolsC}{66}

\newcommand\remove[1]{}

\newcommand{\R}{\mathbb{R}}

\newcommand{\e}{\varepsilon}

\newcommand{\lca}{\mathrm{\bf lca}}
\newcommand{\E}{\mathbb{E}}

\DeclareMathOperator{\diam}{diam}

\newtheorem{theorem}{Theorem}[section]
\newtheorem{lemma}[theorem]{Lemma}
\newtheorem{prop}[theorem]{Proposition}
\newtheorem{claim}[theorem]{Claim}

\theoremstyle{definition}

\newtheorem{observation}[theorem]{Observation}

\newtheorem{remark}{Remark}[section]

\begin{document}

\title{Ramsey partitions and proximity data structures}

\author{ Manor Mendel \and Assaf
Naor}

\date{}
\maketitle

\begin{abstract}

This paper addresses two problems lying at the intersection of
geometric analysis and theoretical computer science: The
non-linear isomorphic Dvoretzky theorem and the design of good
approximate distance oracles for large distortion. We introduce
the notion of Ramsey partitions of a finite metric space, and show
that the existence of good Ramsey partitions implies a solution to
the metric Ramsey problem for large distortion (a.k.a. the
non-linear version of the isomorphic Dvoretzky theorem, as
introduced by Bourgain, Figiel, and Milman in~\cite{BFM86}). We
then proceed to construct optimal Ramsey partitions, and use them
to show that for every $\e\in (0,1)$, every $n$-point metric space
has a subset of size $n^{1-\e}$ which embeds into Hilbert space
with distortion $O(1/\e)$. This result is best possible and
improves part of the metric Ramsey theorem of Bartal, Linial,
Mendel and Naor~\cite{BLMN05}, in addition to considerably
simplifying its proof. We use our new Ramsey partitions to design
approximate distance oracles with a universal constant query time,
closing a gap left open by Thorup and Zwick in~\cite{TZ05}.
Namely, we show that for every $n$ point metric space $X$, and
$k\geq 1$, there exists an $O(k)$-approximate distance oracle
whose storage requirement is $O\left( n^{1+1/k}\right)$, and whose
query time is a universal constant. We also discuss applications
of Ramsey partitions to various other geometric data structure
problems, such as the design of efficient data structures for
approximate ranking.
\end{abstract}

\section{Introduction}

Motivated by the search for a non-linear version of Dvoretzky's
theorem, Bourgain, Figiel and Milman~\cite{BFM86} posed the
following problem, which is known today as the {\em metric Ramsey
problem}: Given a target distortion $\alpha>1$ and an integer $n$,
what is the largest $k$ such that {\em every} $n$-point metric
space has a subset of size $k$ which embeds into Hilbert space
with distortion $\alpha$? (Recall that a metric space $(X,d_X)$ is
said to embed into Hilbert space with distortion $\alpha$ if there
exists a mapping $f:X\to L_2$ such that for every $x,y\in X$, we
have $d_X(x,y)\le \|f(x)-f(y)\|_2\le\alpha d_X(x,y)$). This
problem has since been investigated by several authors, motivated
in part by the discovery of its applications to online algorithms
--- we refer to~\cite{BLMN05} for a discussion of the history and
applications of the metric Ramsey problem.

The most recent work on the metric Ramsey problem is due to
Bartal, Linial, Mendel and Naor~\cite{BLMN05}, who obtained
various nearly optimal upper and lower bounds in several contexts.
Among the results in~\cite{BLMN05} is the following theorem which
deals with the case of large distortion: For every $\e\in (0,1)$,
any $n$-point metric space has a subset of size $n^{1-\e}$ which
embeds into an ultrametric with distortion $O\bigl(\frac{\log
(2/\e)}{\e}\bigr)$ (recall that an ultrametric $(X,d_X)$ is a
metric space satisfying for every $x,y,z\in X$, $d_X(x,y)\le
\max\left\{d_X(x,z),d_X(y,z)\right\}$). Since ultrametrics embed
isometrically into Hilbert space, this is indeed a metric Ramsey
theorem. Moreover, it was shown in~\cite{BLMN05} that this result
is optimal up to the $\log(2/\e)$ factor, i.e. there exists
arbitrarily large $n$-point metric spaces, every subset of which
of size $n^{1-\e}$ incurs distortion $\Omega(1/\e)$ in any
embedding into Hilbert space. The main result of this paper closes
this gap:

\begin{theorem}\label{thm:main}
Let $(X,d_X)$ be an $n$-point metric space and $\e\in (0,1)$. Then
there exists a subset $Y\subseteq X$ with $|Y|\ge n^{1-\e}$ such
that $(Y,d_X)$ is equivalent to an ultrametric with distortion at
most $\frac{128}{\e}$.
\end{theorem}

In the four years that elapsed since our work on~\cite{BLMN05}
there has been remarkable development in the structure theory of
finite metric spaces. In particular, the theory of random
partitions of metric spaces has been considerably refined, and was
shown to have numerous applications in mathematics and computer
science (see for example~\cite{FRT04,LN05-via,KLMN05,ALN05} and
the references therein). The starting point of the present paper
was our attempt to revisit the metric Ramsey problem using random
partitions. It turns out that this approach can indeed be used to
resolve the metric Ramsey problem for large distortion, though it
requires the introduction of a new kind of random partition, an
improved ``padding inequality" for known partitions, and a novel
application of the random partition method in the setting of
Ramsey problems. In Section~\ref{section:part} we introduce the
notion of Ramsey partitions,
 and show how they can be used to
address the metric Ramsey problem. We then proceed in
Section~\ref{section:construct} to construct optimal Ramsey
partitions, yielding Theorem~\ref{thm:main}. Our construction is
inspired in part by Bartal's probabilistic embedding into
trees~\cite{Bartal96}, and is based on a random partition due to
Calinescu, Karloff and Rabani~\cite{CKR04}, with an improved
analysis which strengthens the work of Fakcharoenphol, Rao and
Talwar~\cite{FRT04}.  In particular, our proof of
Theorem~\ref{thm:main} is self contained, and considerably simpler
than the proof of the result from~\cite{BLMN05} quoted above.
Nevertheless, the construction of~\cite{BLMN05} is deterministic,
while our proof of Theorem~\ref{thm:main} is probabilistic.
Moreover, we do not see a simple way to use our new approach to
simplify the proof of another
 main result of~\cite{BLMN05}, namely the phase transition at
distortion $\alpha=2$ (we refer to~\cite{BLMN05} for details, as
this result will not be used here). The results of~\cite{BLMN05}
which were used crucially in our work~\cite{MN04} on the metric
version of Milman's Quotient of Subspace theorem are also not
covered by the present paper.

\bigskip

\noindent{\bf Algorithmic applications to the construction of
proximity data structures.} The main algorithmic application of the
metric Ramsey theorem in~\cite{BLMN05} is to obtain the best known
lower bounds on the competitive ratio of
the randomized $k$-server problem. We refer
to~\cite{BLMN05} and the references therein for more information on
this topic, as Theorem~\ref{thm:main} does not yield improved
$k$-server lower bounds. However, Ramsey partitions are useful to
obtain positive results, and not only algorithmic lower bounds,
which we now describe.

A finite metric space can be thought of as given by its $n\times
n$ distance matrix. However, in many algorithmic contexts it is
worthwhile to preprocess this data so that we store significantly
less than $n^2$ numbers, and still be able to quickly find out
{\em approximately}  the distance between two query points. In
other words, quoting Thorup and Zwick~\cite{TZ05}, ``\emph{In most
applications we are not really interested in {\em all} distances,
we just want the ability to retrieve them quickly, if needed}".
The need for such ``compact" representation of metrics also occurs
naturally in mathematics; for example the methods developed in
theroetical computer science (specifically~\cite{CK95,HM05}) are a
key tool in the recent work of Fefferman and Klartag~\cite{FK05}
on the extension of $C^m$ functions defined on $n$ points in
$\R^d$ to all of $\R^d$.

%Arguably the most

An influential compact representation of metrics used in
theoretical computer science is the {\em approximate distance
oracle}~\cite{ABCP99,Cohen99,TZ05,HM05}. Stated formally, a
$(P,S,Q,D)$-approximate distance oracle on a finite metric space
$(X,d_X)$ is a  data structure that takes expected time $P$ to
preprocess from the given distance matrix, takes space $S$ to
store, and given two query points $x,y\in X$, computes in time $Q$
a number $E(x,y)$ satisfying $d_X(x,y)\le E(x,y)\le D\cdot
d_X(x,y)$. Thus the distance matrix itself is a
$(P=O(1),S=O(n^2),Q=O(1),D=1)$- approximate distance oracle, but
clearly the interest is in {\em compact} data structures in the
sense that $S=o(n^2)$. In what follows we will depart from the
above somewhat cumbersome terminology, and simply discuss
$D$-approximate distance oracles (emphasizing the distortion $D$),
and state in words the values of the other relevant parameters
(namely the preprocessing time, storage space and query time).

An important paper of Thorup and Zwick~\cite{TZ05} constructs the
best known approximate distance oracles. Namely, they show that
for every integer $k$, every $n$-point metric space has a
$(2k-1)$-approximate distance oracle which can be preprocessed in
time $O\left(n^{2}\right)$, requires storage $O\left(k\cdot
n^{1+1/k}\right)$, and has query time $O(k)$. Moreover, it is
shown in~\cite{TZ05} that this distortion/storage tradeoff is
almost tight: A widely believed combinatorial conjecture of
Erd\H{o}s~\cite{Erd64} is shown in~\cite{TZ05} (see
also~\cite{Mat96}) to imply that any data structure supporting
approximate distance queries with distortion at most $2k-1$ must
be of size at least $\Omega\left(n^{1+1/k}\right)$ bits. Since for
large values of $k$ the query time of the Thorup-Zwick oracle is
large,  the problem remained whether there exist good approximate
distance oracles whose query time is a constant independent of the
distortion (i.e., in a sense, true ``oracles").  Here we use
Ramsey partitions to answer this question positively: For any
distortion, every metric space admits an approximate distance
oracle with storage space almost as good as the Thorup-Zwick
oracle (in fact, for distortions larger than $\Omega(\log n/ \log
\log n)$ our storage space is slightly better), but whose query
time is a universal constant. Stated formally, we prove the
following theorem:

\begin{theorem}\label{thm:oracle} For any $k>1$, every $n$-point metric space
$(X,d_X)$ admits a $O(k)$-approximate distance oracle whose
preprocessing time is $O\left(n^{2+1/k}\log n\right)$, requiring
storage space $O\left( n^{1+1/k}\right)$, and whose query time is a
universal constant.
\end{theorem}

Another application of Ramsey partitions is to the construction of
data structures for {\em approximate ranking}. This problem is
motivated in part by web search and the analysis of social networks,
in addition to being a natural extension of the ubiquitous
approximate nearest neighbor search problem
(see~\cite{AMNSW98,Ind04-handbook,Clarkson05} and the references therein).
In the
approximate nearest neighbor search problem we are given $c>1$, a
metric space $(X,d_X)$, and a subset $Y\subseteq X$. The goal is to
preprocess the data points $Y$ so that given a query point $x\in
X\setminus Y$ we quickly return a point $y\in Y$ which is a
$c$-approximate nearest neighbor of $x$, i.e. $d_X(x,y)\le
cd_X(x,Y)$. More generally, one might want to find the second
closest point to $x$ in $Y$, and so forth
(this problem has been studied extensively in computational geometry,
see for example~\cite{AMNSW98}). In other words, by
ordering the points in $X$ in increasing distance from $x\in X$ we
induce a {\em proximity ranking} of the points of $X$. Each point of
$X$ induces a different ranking of this type, and computing it
efficiently is a natural generalization of the nearest neighbor
problem. Using our new Ramsey partitions we design the following
data structure for solving this problem approximately:

\begin{theorem}\label{thm:ranking}
Fix $k>1$, and an $n$-point metric space $(X,d_X)$. Then there exist
a data structure which can be preprocessed in time $O\left(k
n^{2+1/k}\log n\right)$, uses only $O\left(k n^{1+1/k}\right)$
storage space, and supports the following type of queries: Given
$x\in X$, have ``fast access" to a permutation of $\pi^{(x)}$ of $X$
satisfying for every $1\le i<j\le n$,
$d_X\left(x,\pi^{(x)}(i)\right)\leq O(k) \cdot
d_X\left(x,\pi^{(x)}(j)\right)$. By ``fast access" to $\pi^{(x)}$ we
mean that we can do the following:
\begin{compactenum}
\item Given a point $x\in X$, and $i\in \{1,\ldots,n\}$, find $\pi^{(x)}(i)$ in constant time.
\item For any $x,u\in X$,  compute $j \in \{1,\ldots,n\}$ such that $\pi^{(x)}(j)=u$ in
constant time.
\end{compactenum}
\end{theorem}

\bigskip

As is clear from the above discussion, the present paper is a
combination of results in pure mathematics, as well as the theory
of data structures. This exemplifies the close interplay between
geometry and computer science, which has become a major driving
force in modern research in these areas. Thus, this paper
``caters" to two different communities, and we put effort into
making it accessible to both.

\begin{comment}
The close interplay between geometry and computer science has
become a major driving force in modern research in these areas. We
hope that our work will contribute to enhancing this beautiful and
fruitful ``cooperation" in the future.
\end{comment}

\section{Ramsey partitions and their equivalence to the metric Ramsey
problem}\label{section:part}

Let $(X,d_X)$ be a metric space. In what follows for $x\in X$ and
$r\ge 0$ we let $B_X(x,r)=\{y\in X:\ d_X(x,y)\le r\}$ be the {\em
closed} ball of radius $r$ centered at $x$. Given a partition
$\mathscr P$ of $X$ and $x\in X$ we denote by $\mathscr P(x)$ the
unique element of $\mathscr P$ containing $x$. For $\Delta>0$ we
say that $\mathscr P$ is $\Delta$-bounded if for every $C\in
\mathscr P$, $\diam(C)\le \Delta$. A {\em partition tree} of $X$
is a sequence of partitions $\{\mathscr P_k\}_{k=0}^\infty$ of $X$
such that $\mathscr P_0=\{X\}$, for all $k\ge 0$ the partition
$\mathscr P_k$ is $ 8^{-k}\diam(X)$-bounded, and $\mathscr
P_{k+1}$ is a refinement of $\mathscr P_k$ (the choice of $8$ as
the base of the exponent in this definition is convenient, but
does not play a crucial role here). For $\beta,\gamma>0$ we shall
say that a probability distribution $\Pr$ over partition trees
$\{\mathscr {P}_k\}_{k=0}^\infty$ of $X$ is completely
$\beta$-padded with exponent $\gamma$ if for every $x\in X$,
$$
\Pr\left[\forall\ k\in \mathbb N,\ B_X\left(x,\beta\cdot
8^{-k}\diam(X)\right)\subseteq \mathscr P_k(x)\right]\ge
|X|^{-\gamma}.
$$
We shall call such probability distributions over partition trees
{\em Ramsey partitions}.

The following lemma shows that the existence of good Ramsey
partitions implies a solution to the metric Ramsey problem. In
fact, it is possible to prove the converse direction, i.e. that
the metric Ramsey theorem implies the existence of good Ramsey
partitions (with appropriate dependence on the various
parameters). We defer the proof of this implication to
Appendix~\ref{app:converse} as it will not be used in this paper
due to the fact that in Section~\ref{section:construct} we will
construct directly optimal Ramsey partitions.

\begin{lemma}\label{lem:ramseypart} Let $(X,d_X)$ be an $n$-point metric space which
admits a distribution over partition trees which is completely
$\beta$-padded with exponent $\gamma$. Then there exists a subset
$Y\subseteq X$ with $|Y|\ge n^{1-\gamma}$ which is $8/\beta$
equivalent\footnote{Here, and in what follows, for $D\ge 1$ we say
that two metric spaces $(X,d_X)$ and $(Y,d_Y)$ are $D$-equivalent
if there exists a bijection $f:X\to Y$ and a scaling factor $C>0$
such that for all $x,y\in X$ we have $Cd_X(x,y)\le
d_Y(f(x),f(y))\le CDd_X(x,y)$.} to an ultrametric.
\end{lemma}

\begin{proof} We may assume without loss of generality that
$\diam(X)=1$. Let $\{\mathscr P_k\}_{k=0}^\infty$ be a distribution
over partition trees of $X$ which is completely $\beta$-padded with
exponent $\gamma$. We define an ultrametric $\rho$ on $X$ as
follows. For $x,y\in X$ let $k$ be the largest integer for which
$\mathscr P_k(x)=\mathscr P_k(y)$, and set $\rho(x,y)=8^{-k}$. It is
straightforward to check that $\rho$ is indeed an ultrametric.
Consider the random subset $Y\subseteq X$ given by
$$
Y=\left\{x\in X:\ \forall\ k\in \mathbb N,\ B_X\left(x,\beta\cdot
8^{-k}\right)\subseteq \mathscr P_k(x)\right\}.
$$
Then
$$
\E |Y|=\sum_{x\in X}\Pr\left[\forall\ k\in \mathbb N,\
B_X\left(x,\beta\cdot 8^{-k}\diam(X)\right)\subseteq \mathscr
P_k(x)\right]\ge n^{1-\gamma}.
$$
We can therefore choose $Y\subseteq X$ with $|Y|\ge n^{1-\gamma}$
such that for all $x\in Y$ and all $k\ge 0$ we have
$B_X\left(x,\beta\cdot 8^{-k}\right)\subseteq \mathscr P_k(x)$.
Fix $x,y\in X$, and let $k$ be the largest integer for which
$\mathscr P_k(x)=\mathscr P_k(y)$. Then $d_X(x,y)\le
\diam(\mathscr P_k(x))\le 8^{-k}=\rho(x,y)$. On the other hand, if
$x\in X$ and $y\in Y$ then, since $\mathscr P_{k+1}(x)\neq
\mathscr P_{k+1}(y)$, the choice of $Y$ implies that $x\notin
B_X\left(y,\beta\cdot 8^{-k-1}\right)$. Thus $d_X(x,y)> \beta\cdot
8^{-k-1}=\frac{\beta}{8}\rho(x,y)$. It follows that the metrics
$d_X$ and $\rho$ are equivalent on $Y$ with distortion $8/\beta$.
\end{proof}

\section{Constructing optimal Ramsey
partitions}\label{section:construct}

The following lemma gives improved bounds on the ``padding
probability" of a distribution over partitions which was discovered
by Calinescu, Karloff and Rabani in~\cite{CKR04}.

\begin{lemma}\label{lem:improvedCKR} Let $(X,d_X)$ be a finite
metric space. Then for every $\Delta>0$ there exists a probability
distribution $\Pr$ over $\Delta$-bounded partitions of $X$ such
that for every $0<t\le\Delta/8$ and every $x\in X$,
\begin{eqnarray}
\label{eq:betterCKR} \Pr\left[B_X\left(x,t\right)\subseteq \mathscr
P(x)\right]\ge
\left(\frac{\left|B_X(x,\Delta/8)\right|}{\left|B_X(x,\Delta)\right|}\right)^{\frac{16t}{\Delta}}.
\end{eqnarray}
\end{lemma}

\begin{remark} {The distribution over partitions used in the proof of
Lemma~\ref{lem:improvedCKR} is precisely the distribution introduced
by Calinescu, Karloff and Rabani in~\cite{CKR04}. In~\cite{FRT04}
Fakcharoenphol, Rao and Talwar proved the following estimate for the
same distribution
\begin{eqnarray}\label{eq:FRT}
\Pr\left[B_X\left(x,t\right)\subseteq \mathscr
P(x)\right]\ge1-O\left(\frac{t}{\Delta}\log
\frac{\left|B_X(x,\Delta)\right|}{\left|B_X(x,\Delta/8)\right|}\right).
\end{eqnarray}
Clearly the bound~\eqref{eq:betterCKR} is stronger
than~\eqref{eq:FRT}, and in particular it yields a non-trivial
estimate even for large values of $t$ for which the lower bound
in~\eqref{eq:FRT} is negative. This improvement is crucial for our
proof of Theorem~\ref{thm:main}. The use of the ``local ratio of
balls" (or ``local growth") in the estimate~\eqref{eq:FRT} of
Fakcharoenphol, Rao and Talwar was a fundamental breakthrough,
which, apart from their striking application in~\cite{FRT04}, has
since found several applications in mathematics and computer
science (see~\cite{LN05-via,KLMN05,ALN05}).}
\end{remark}

\begin{proof}[Proof of Lemma~\ref{lem:improvedCKR}] Write $X=\{x_1,\ldots,x_n\}$. Let $R$ be chosen
uniformly at random from the interval $[\Delta/4,\Delta/2]$, and let
$\pi$ be a permutation of $\{1,\ldots,n\}$ chosen uniformly at
random from all such permutations (here, and in what follows, $R$
and $\pi$ are independent). Define $C_1\coloneqq
B_X\left(x_{\pi(1)},R\right)$ and inductively for $2\le j\le n$,
$$
C_j\coloneqq B_X\left(x_{\pi(j)},R\right)\setminus
\bigcup_{i=1}^{j-1}C_i.
$$
Finally we let $\mathscr P\coloneqq \{C_1,\ldots,C_n\}\setminus
\{\emptyset\}$. Clearly $\mathscr P$ is a (random) $\Delta$-bounded
partition on $X$.

For every $r\in [\Delta/4,\Delta/2]$,
\begin{equation}\label{eq:ratio}
\Pr\left[B_X\left(x,t\right)\subseteq \mathscr P(x) |R=r\right] \ge
\frac{\left|B_X(x,r-t)\right|}{\left|B_X(x,r+t)\right|}.
\end{equation}
Indeed, if $R=r$, then the triangle inequality implies that if in
the random order induced by the partition $\pi$ on the points of
the ball $B_X(x,r+t)$ the minimal element is from the ball
$B_X(x,r-t)$, then $B_X\left(x,t\right)\subseteq \mathscr P(x)$
(see Figure~\ref{fig:separating} for a schematic description of
this situation). This event happens with probability
$\frac{\left|B_X(x,r-t)\right|}{\left|B_X(x,r+t)\right|}$,
implying~\eqref{eq:ratio}.

\begin{figure}[h]
\begin{center} \includegraphics[scale=0.78]{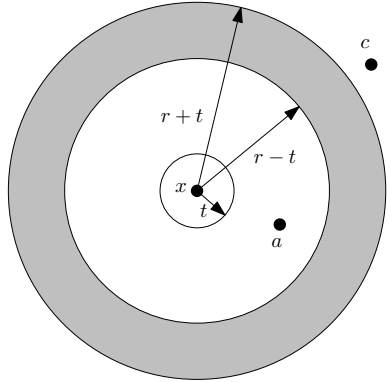} \end{center}
 \caption{{\small{\em  A schematic description of the lower bound
 in~\eqref{eq:ratio}}. The clusters that
 are induced by points which lie outside the
 ball $B_X(x,r+t)$, such as $c$, cannot touch the ball $B_X(x,t)$. On
 the other hand, if a point from $B_X(x,r-t)$, such as $a$, appeared first in the
 random order among the points in $B_X(x,r+t)$ then its cluster
 will ``swallow" the ball $B_X(x,t)$. The probability for this to happen is
$\frac{|B(x,r-t)|}{|B(x,r+t)|}$. Only points in the shaded
 region can split the ball $B_X(x,t)$.}} \label{fig:separating}
\end{figure}

Write $\frac{\Delta}{8t}=k+\beta$, where $\beta\in [0,1)$ and $k$
is a positive integer. Then
\begin{eqnarray}
\label{eq:use condition}\Pr\left[B_X\left(x,t\right)\subseteq
\mathscr P(x)\right]&\ge& \frac{4}{\Delta}\int_{\Delta/4}^{\Delta/2}
\frac{\left|B_X(x,r-t)\right|}{\left|B_X(x,r+t)\right|}dr\\
&=& \nonumber\frac{4}{\Delta}\sum_{j=0}^{k-1
}\int_{\frac{\Delta}{4}+2jt}^{\frac{\Delta}{4}+2(j+1)t}\frac{\left|B_X(x,r-t)\right|}{\left|B_X(x,r+t)\right|}dr+\frac{4}{\Delta}
\int_{\frac{\Delta}{4}+2kt}^{\frac{\Delta}{2}}\frac{\left|B_X(x,r-t)\right|}{\left|B_X(x,r+t)\right|}dr\\
&\ge& \nonumber \frac{4}{\Delta}\int_0^{2t}\sum_{j=0}^{k-1
}\frac{\left|B_X\left(x,\frac{\Delta}{4}+2jt+s-t\right)\right|}{\left|B_X\left(x,\frac{\Delta}{4}+2jt+s+t\right)\right|}ds+
\frac{4}{\Delta}\left(\frac{\Delta}{4}-2kt\right)\frac{\left|B_X\left(x,\frac{\Delta}{4}+2kt-t\right)\right|}{\left|B_X\left(x,\frac{\Delta}{2}+t\right)\right|}\\
&\ge&\label{eq:AMGM} \frac{4k}{\Delta}\int_0^{2t}
\left[\prod_{j=0}^{k-1
}\frac{\left|B_X\left(x,\frac{\Delta}{4}+2jt+s-t\right)\right|}{\left|B_X\left(x,\frac{\Delta}{4}+2jt+s+t\right)\right|}
\right]^{\frac{1}{k}}ds+\left(1-\frac{8kt}{\Delta}\right)\frac{\left|B_X\left(x,\frac{\Delta}{4}+2kt-t\right)\right|}{\left|B_X\left(x,\frac{\Delta}{2}+t\right)\right|}\\
&=& \nonumber \frac{4k}{\Delta}\cdot \int_0^{2t}
\left[\frac{\left|B_X\left(x,\frac{\Delta}{4}+s-t\right)\right|}{\left|B_X\left(x,\frac{\Delta}{4}+2t\left(k-1\right)+s+t\right)\right|}
\right]^{\frac{1}{k}}ds+\left(1-\frac{8kt}{\Delta}\right)\frac{\left|B_X\left(x,\frac{\Delta}{4}+2kt-t\right)\right|}{\left|B_X\left(x,\frac{\Delta}{2}+t\right)\right|}\\
&\ge& \nonumber\frac{8kt}{\Delta}
\left[\frac{\left|B_X\left(x,\frac{\Delta}{4}-t\right)\right|}
{\left|B_X\left(x,\frac{\Delta}{4}+2kt+t\right)\right|}
\right]^{\frac{1}{k}}+\left(1-\frac{8kt}{\Delta}\right)\frac{\left|B_X\left(x,\frac{\Delta}{4}+2kt-t\right)\right|}{\left|B_X\left(x,\frac{\Delta}{2}+t\right)\right|}\\
&\ge& \label{eq:log concave}
\left[\frac{\left|B_X\left(x,\frac{\Delta}{4}-t\right)\right|}
{\left|B_X\left(x,\frac{\Delta}{4}+2kt+t\right)\right|}\right]^{\frac{8t}{\Delta}}\cdot
\left[\frac{\left|B_X\left(x,\frac{\Delta}{4}+2kt-t\right)\right|}{\left|B_X\left(x,\frac{\Delta}{2}+t\right)\right|}\right]^{1-\frac{8kt}{\Delta}}\\
&=& \nonumber
\left[\frac{\left|B_X\left(x,\frac{\Delta}{4}-t\right)\right|}
{\left|B_X\left(x,\frac{\Delta}{4}+2kt+t\right)\right|}\cdot
\frac{\left|B_X\left(x,\frac{\Delta}{4}+2kt-t\right)\right|}{\left|B_X\left(x,\frac{\Delta}{2}+t\right)\right|}
\right]^{\frac{8t}{\Delta}}\cdot
\left[\frac{\left|B_X\left(x,\frac{\Delta}{4}+2kt-t\right)\right|}{\left|B_X\left(x,\frac{\Delta}{2}+t\right)\right|}\right]
^{\frac{8t}{\Delta}\left(\frac{\Delta}{8t}-k-1\right)}\\
&\ge& \label{eq:negative}
\left[\frac{\left|B_X\left(x,\frac{\Delta}{4}-t\right)\right|}
{\left|B_X\left(x,\frac{\Delta}{2}+t\right)\right|}
\right]^{\frac{16t}{\Delta}},
\end{eqnarray}
where in~\eqref{eq:use condition} we used~\eqref{eq:ratio},
in~\eqref{eq:AMGM} we used the arithmetic mean/geometric mean
inequality, in~\eqref{eq:log concave} we used the elementary
inequality $\theta a+(1-\theta)b\ge a^\theta b^{1-\theta}$, which
holds for all $\theta \in [0,1]$ and $a,b\ge 0$, and
in~\eqref{eq:negative} we used the fact that $\frac{\Delta}{8t}-k-1$
is negative.
\end{proof}

% Y. Bartal informed us that a variant of his random partitions~\cite{Bartal05}
% can also be used
% to obtain a result similar to Lemma~\ref{lem:improvedCKR}.

The following theorem, in conjunction with
Lemma~\ref{lem:ramseypart}, implies Theorem~\ref{thm:main}.

\begin{theorem}\label{thm:part-tree} For every $\alpha>1$, every finite metric space $(X,d_X)$
admits a completely $1/\alpha$ padded random partition tree with
exponent $16/\alpha$.
\end{theorem}

\begin{proof}
Fix $\alpha>1$. Without loss of generality we may assume that
$\diam(X)=1$. We construct a partition tree  $\{\mathscr
E_k\}_{k=0}^\infty$ of $X$ as follows. Set $\mathscr E_0=\{X\}$.
Having defined $\mathscr E_k$ we let $\mathscr P_{k+1}$ be a
partition as in Lemma~\ref{lem:improvedCKR} with $\Delta=8^{-k}$ and
$t=\Delta/\alpha$ (the random partition $\mathscr P_{k+1}$ is chosen
independently of the random partitions $\mathscr P_1,\ldots,\mathscr
P_k$). Define $\mathscr E_{k+1}$ to be the common refinement of
$\mathscr E_k$ and $\mathscr P_{k+1}$, i.e.
$$
\mathscr E_{k+1}\coloneqq \{C\cap C':\ C\in \mathscr E_k,\ C'\in
\mathscr P_{k+1}\}.
$$
The construction implies that for every $x\in X$ and every $k\ge 0$
we have $\mathscr E_{k+1}(x)=\mathscr E_k(x)\cap \mathscr
P_{k+1}(x)$. Thus one proves inductively that
$$ \forall\  k\in
\mathbb N,\ B_X\left(x,\frac{8^{-k}}{\alpha}\right)\subseteq
\mathscr P_{k}(x)\implies \forall\  k\in \mathbb N,\
B_X\left(x,\frac{8^{-k}}{\alpha}\right)\subseteq \mathscr E_{k}(x).
$$
From Lemma~\ref{lem:improvedCKR} and the independence of $\{\mathscr
P_k\}_{k=1}^\infty$ it follows that
\begin{eqnarray*}
\Pr\left[\forall\  k\in \mathbb N,\
B_X\left(x,\frac{8^{-k}}{\alpha}\right)\subseteq \mathscr
E_{k}(x)\right]&\ge& \Pr\left[\forall\  k\in \mathbb N,\
B_X\left(x,\frac{8^{-k}}{\alpha}\right)\subseteq \mathscr
P_{k}(x)\right]\\
&=&\prod_{k=1}^\infty\Pr\left[
B_X\left(x,\frac{8^{-k}}{\alpha}\right)\subseteq \mathscr
P_{k}(x)\right]\\
&\ge&
\prod_{k=1}^\infty\left[\frac{|B_X(x,8^{-k-1})|}{|B_X(x,8^{-k})|}\right]^{\frac{16}{\alpha}}\\
&=& |B_X(x,1/8)|^{-\frac{16}{\alpha}}\ge |X|^{-\frac{16}{\alpha}}.
\qedhere
\end{eqnarray*}
\end{proof}

%\section{An application to approximate ranking data
%structures}\label{section:use}

\section{Applications to proximity data structures}\label{section:use}

%\paragraph{Approximate Distance Oracles}

In this section we show how Theorem~\ref{thm:part-tree} can be
applied to the design of various proximity data structures, which
are listed below. Before doing so we shall recall some standard
facts about tree representations of ultrametrics, all of which can
be found in the discussion in~\cite{BLMN05}. Any finite
ultrametric $(X,\rho)$ can be represented by a rooted tree
$T=(V,E)$ with labels $\Delta:V\to (0,\infty)$, whose leaves are
$X$, and such that if $u,v\in V$ and $v$ is a child of $u$ then
$\Delta(v)\le \Delta(u)$. Given $x,y\in X$ we then have
$\rho(x,y)=\Delta\left(\lca(x,y)\right)$, where $\lca(x,y)$ is the
least common ancestor of $x$ and $y$ in $T$. For $k\ge 1$ the
labelled tree described above is called a $k$-HST (hierarchically
well separated tree) if its labels satisfy the stronger decay
condition $\Delta(v)\le \frac{\Delta(u)}{k}$ whenever $v$ is a
child of $u$. The tree $T$ is called an exact $k$-HST if we
actually have an equality $\Delta(v)= \frac{\Delta(u)}{k}$
whenever $v$ is a child of $u$. Lemma 3.5 in~\cite{BLMN05} implies
that any $n$-point ultrametric is $k$-equivalent to a metric on
$k$-HST which can be computed in time $O(n)$.

We start by proving several structural lemmas which will play a
crucial role in the design of our new data structures.

\begin{lemma}[Extending ultrametrics]\label{lem:extend} Let
$(X,d_X)$ be a finite metric space, and $\alpha\ge 1$. Fix
$\emptyset\neq Y\subseteq X$, and assume that there exits an
ultrametric $\rho$ on $Y$ such that for every $x,y\in Y$,
$d_X(x,y)\le \rho(x,y)\le \alpha d_X(x,y)$. Then there exists an
ultrametric $\widetilde \rho$ defined on all of $X$ such that for
every $x,y\in X$ we have $d_X(x,y)\le \widetilde \rho(x,y)$, and
if $x\in X$ and $y\in Y$ then $\widetilde \rho(x,y)\le 6\alpha
d_X(x,y)$.
\end{lemma}

\begin{proof} Let $T=(V,E)$ be the 1-HST representation of $\rho$,
with labels $\Delta:V\to (0,\infty)$. In other words, the leaves
of $T$ are $Y$, and for every $x,y\in Y$ we have
$\Delta(\lca(x,y))=\rho(x,y)$. It will be convenient to augment
$T$ by adding an incoming edge to the root with
$\Delta(\mathrm{parent}(\mathrm{root}))=\infty$. This clearly does
not change the induced metric on $Y$. For every $x\in X\setminus
Y$ let $y\in Y$ be its closest point in $Y$, i.e.
$d_X(x,y)=d_X(x,Y)$. Let $u$ be the least ancestor of $y$ for
which $\Delta(u)\ge d_X(x,y)$ (such a $u$ must exist because we
added the incoming edge to the root). Let $v$ be the child of $u$
along the path connecting $u$ and $y$. We add a vertex $w$ on the
edge $\{u,v\}$ whose label is $d_X(x,y)$, and connect $x$ to $T$
as a child of $w$. The resulting tree is clearly still a $1$-HST.
Repeating this procedure for every $x\in X\setminus Y$ we obtain a
$1$-HST $\widetilde T$ whose leaves are $X$. Denote the labels on
$\widetilde T$ by $\widetilde \Delta$.

Fix $x,y\in X$, and let $x',y'\in Y$ the nearest neighbors of
$x,y$ (respectively) used in the above construction. Then
\begin{eqnarray}\label{eq:lowerbound}
\widetilde \Delta\left(\lca_{\widetilde T}(x,y)\right)&=&\nonumber
\max\left\{\widetilde \Delta\left(\lca_{\widetilde
T}(x,x')\right),\widetilde \Delta\left(\lca_{\widetilde
T}(y,y')\right),\widetilde \Delta\left(\lca_{\widetilde
T}(x',y')\right)\right\}\nonumber\\&\ge&
\max\left\{d_X(x,x'),d_X(y,y'),d_X(x',y')\right\}\nonumber\\&\ge&
\frac{d_X(x,x')+d_X(y,y')+d_X(x',y')}{3}\nonumber\\
&\ge& \frac13\, d_X(x,y).
\end{eqnarray}

In the reverse direction, if $x\in X$ and $y\in Y$ let $x'\in Y$
be the closest point in $Y$ to $x$ used in the construction of
$\widetilde T$. Then $d_X(x',y)\le d_X(x',x)+d_X(x,y)\le
2d_X(x,y)$. If $\lca_{\widetilde T}(y,x')$ is an ancestor of
$\lca_{\widetilde T}(x,x')$ then
\begin{eqnarray}\label{eq:upperbound1}
\widetilde \Delta\left(\lca_{\widetilde
T}(x,y)\right)=\widetilde\Delta\left(\lca_{\widetilde
T}(x',y)\right)=\rho(x',y)\le \alpha\cdot d_X(x',y)\le
2\alpha\cdot d_X(x,y). \end{eqnarray} If, on the other hand,
$\lca_{\widetilde T}(y,x')$ is a descendant of $\lca_{\widetilde
T}(x,x')$ then
\begin{eqnarray}\label{eq:upperbound2}
\widetilde \Delta\left(\lca_{\widetilde
T}(x,y)\right)=\widetilde\Delta\left(\lca_{\widetilde
T}(x,x')\right)= d_X(x,x')\le d_X(x,y).
\end{eqnarray}
Scaling the labels of $\widetilde T$ by a factor of $3$, the
required result is a combination of~\eqref{eq:lowerbound},
\eqref{eq:upperbound1} and~\eqref{eq:upperbound2}.
\end{proof}

The following lemma is a structural result on the existence of a
certain distribution over decreasing chains of subsets of a finite
metric space. In what follows we shall call such a distribution a
{\em stochastic Ramsey chain}. A schematic description of this
notion, and the way it is used in the ensuing arguments, is
presented in Figure~\ref{fig:oracles} below.

\begin{lemma}[Stochastic Ramsey chains]\label{lem:sample} Let $(X,d_X)$ be an $n$-point metric
space and $k\ge 1$. Then there exists a distribution over
decreasing sequences of subsets $X=X_0\supsetneqq X_1\supsetneqq
X_2\cdots \supsetneqq X_s=\emptyset$ ($s$ itself is a random
variable), such that for all $p> -1/k$,
\begin{eqnarray}\label{eq:moments}
\E\left[\sum_{j=0}^{s-1} |X_j|^p\right]\le
\left(\max\left\{\frac{k}{1+pk},1\right\}\right)\cdot n^{p+1/k},
\end{eqnarray}
and such that for each $j\in \{1,\ldots,s\}$ there exists an
ultrametric $\rho_j$ on $X$ satisfying for every $x,y\in X$,
$\rho_j(x,y)\ge d_X(x,y)$, and if $x\in X$ and $y\in
X_{j-1}\setminus X_{j}$ then $\rho_j(x,y)\le O(k)\cdot d_X(x,y)$.
\end{lemma}

\begin{remark}{In what follows we will only use the cases $p\in
\{0,1,2\}$ in Lemma~\ref{lem:sample}. Observe that for $p=0$,
\eqref{eq:moments} is simply the estimate $\E s\le kn^{1/k}$. }
\end{remark}

\begin{proof}[Proof of Lemma~\ref{lem:sample}]
By Theorem~\ref{thm:part-tree} and the proof of
Lemma~\ref{lem:ramseypart} there is a distribution over subsets
$Y_1\subseteq X_0$ such that $\E|Y_1|\ge n^{1-1/k}$ and there
exists an ultrametric $\rho_1$ on $Y_1$ such that every $x,y\in
Y_1$ satisfy $d_X(x,y)\le \rho_1(x,y)\le O(k)\cdot d_X(x,y)$. By
Lemma~\ref{lem:extend} we may assume that $\rho_1$ is defined on
all of $X$, for every $x,y\in X$ we have $\rho_1(x,y)\ge
d_X(x,y)$, and if $x\in X$ and $y\in Y_1$ then $\rho_1(x,y)\le
O(k)\cdot d_X(x,y)$. Define $X_1=X_0\setminus Y_1$ and apply the
same reasoning to $X_1$, obtaining a
 random
subset $Y_2\subseteq X_0\setminus Y_1$ and an ultrametric
$\rho_2$. Continuing in this manner until we arrive at the empty
set, we see that there are disjoint subsets,
$Y_1,\ldots,Y_s\subseteq X$, and for each $j$ an ultrametric
$\rho_j$ on $X$, such that for $x,y\in X$ we have $\rho_j(x,y)\ge
d_X(x,y)$, and for $x\in X$, $y\in Y_j$ we have $\rho_j(x,y)\le
O(k)\cdot d_X(x,y)$. Additionally, writing $X_{j}\coloneqq
X\setminus \bigcup_{i=1}^{j}Y_i$, we have the estimate $\E
\left[|Y_j|\Big|Y_1,\ldots,Y_{j-1}\right]\ge |X_{j-1}|^{1-1/k}$.

The proof of~\eqref{eq:moments} is by induction on $n$. For $n=1$
the claim is obvious, and if $n>1$ then by the inductive
hypothesis
\begin{eqnarray*}
\E \left[\left.\sum_{j=0}^{s-1}|X_j|^p\right|Y_1\right]&\le&
n^p+\left(\max\left\{\frac{k}{1+pk},1\right\}\right)\cdot|X_1|^{p+1/k}\\&=&n^p+\left(\max\left\{\frac{k}{1+pk},1\right\}\right)\cdot
n^{p+1/k}\left(1-\frac{|Y_1|}{n}\right)^{p+1/k}\\&\le&
n^p+\left(\max\left\{\frac{k}{1+pk},1\right\}\right)\cdot
n^{p+1/k}\left(1-\left(\min\left\{p+\frac{1}{k},1\right\}\right)\cdot\frac{|Y_1|}{n}\right)\\&=&\left(\max\left\{\frac{k}{1+pk},1\right\}\right)\cdot
n^{p+1/k}+n^p-n^{p-1+1/k} |Y_1|.
\end{eqnarray*}
Taking expectation with respect to $Y_1$ gives the required
result.
\end{proof}

\begin{observation}
If one does not mind losing a factor of $O(\log n)$ in the
construction time and storage of the Ramsey chain, then an
alternative to Lemma~\ref{lem:sample} is to randomly and
independently sample $O\left(n^{1/k} \log n\right)$ ultrametrics
from the Ramsey partitions.
\end{observation}

\begin{comment}
\begin{lemma}\label{lem:sample} Let $(X,d_X)$ be an $n$-point metric space. Then there
exists a collection of ultrametrics $\{\rho_j\}_{j=1}^s$ with
$s=O\left(n^{1/k}\log n\right)$, disjoint subsets
$Y_1,\ldots,Y_s\subseteq X$ such that $X=\bigcup_{j=1}^s Y_j$, and
for every $x\in X$ and $y\in Y_j$, $d_X(x,y)\le \rho_j(x,y)\le
O(k)\cdot d_X(x,y)$.
\end{lemma}

\begin{proof} We may assume that $k\ge 16$. By Theorem~\ref{thm:part-tree} there
is a distribution over partition trees $\{\mathscr
P_j\}_{j=0}^\infty$ which is completely $16/k$ padded with
exponent $1/k$. As in the proof Lemma~\ref{lem:ramseypart} we
assume that $\diam(X)=1$ and denote
$$
Y_0=\left\{x\in X:\ \forall\ j\in \mathbb N,\
B_X\left(x,\frac{16}{8^jk } \right)\subseteq \mathscr
P_j(x)\right\}.
$$
As in the proof of Lemma~\ref{lem:ramseypart}, define for $x,y\in
X$, $\rho_m(x,y)=8^{-j}$, where $j$ is the largest integer for which
$\mathscr P_j^{(m)}(x)=\mathscr P_j^{(m)}(y)$. The metrics
$\rho_1,\ldots,\rho_s$ are ultrametrics, and the proof of
Lemma~\ref{lem:ramseypart} shows that if $x\in X$ and $y\in Y_m$
then $d_X(x,y)\le \rho_m(x,y)\le \frac{k}{2}\cdot d_X(x,y)$. Now,
$$
\Pr\left[X=\bigcup_{m=1}^s Y_m\right]\ge 1-\sum_{x\in
X}\Pr\left[\forall\ 1\le m\le s \ x\notin Y_m\right]\ge
1-n\left(1-\frac{1}{n^{1/k}}\right)^s\ge \frac12,
$$
provided that $s\ge 2n^{1/k}\log n$. This shows that the required
assertions hold with constant positive probability.
\end{proof}
\end{comment}

Before passing to the description of our new data structures, we
need to say a few words about the algorithmic implementation of
Lemma~\ref{lem:sample} (this will be the central preprocessing
step in our constructions). The computational model in which we
will be working is the RAM model, which is standard in the context
of our type of data-structure problems (see for
example~\cite{TZ05}). In fact, we can settle for weaker
computational models such as the ``Unit cost floating-point word
RAM model" --- a detailed discussion of these issues can be found
in Section 2.2. of~\cite{HM05}.

The natural implementation of the Calinescu-Karloff-Rabani (CKR)
random partition used in the proof of Lemma~\ref{lem:improvedCKR}
takes $O\left(n^2\right)$ time. Denote by $\Phi=\Phi(X)$ the
aspect ratio of $X$, i.e. the diameter of $X$ divided by the
minimal positive distance in $X$. The construction of the
distribution over partition trees in the proof of
Theorem~\ref{thm:part-tree} requires performing $O(\log \Phi)$
such decompositions. This results in $O\left(n^2 \log \Phi\right)$
preprocessing time to sample one partition tree from the
distribution. Using a standard technique (described for example
in~\cite[Sections~3.2-3.3]{HM05}), we dispense with the dependence
on the aspect ratio and obtain that the expected preprocessing
time of one partition tree is $O\left(n^{2} \log n\right)$. Since
the argument in~\cite{HM05} is presented in a slightly different
context, we shall briefly sketch it here.

We start by constructing an ultrametric $\rho$ on $X$, represented
by an HST $H$, such that for every $x,y\in X$, $d_X(x,y)\le
\rho(x,y)\le nd_X(x,y)$. The fact that such a tree exists is
contained in~\cite[Lemma~3.6]{BLMN05}, and it can be constructed
in time $O\left(n^2\right)$ using the Minimum Spanning Tree
algorithm. This implementation is done in~\cite[Section~3.2]{HM05}.
We then apply the CKR random partition with diameter $\Delta$ as
follows: Instead of applying it to the points in $X$, we apply it
to the vertices $u$ of $H$ for which
\begin{equation}\label{eq:tree condition} \Delta(u)\le
\frac{\Delta}{n^2}<\Delta\left(\mathrm{parent}(u)\right).
\end{equation}
Each such vertex $u$ represents all the subtree rooted at $u$ (in
particular, we can choose arbitrary leaf descendants to calculate
distances --- these distances are calculated using the metric
$d_X$), and they are all assigned to the same cluster as $u$ in the
resulting partition. This is essentially an application of the
algorithm to an appropriate quotient of $X$ (see the discussion
in~\cite{MN04}). We actually apply a weighted version of the CKR
decomposition in the spirit of~\cite{LN05-via}, in which, in the
choice of random permutation, each vertex $u$ as above is chosen
with probability proportional to the number of leaves which are
descendants of $u$ (note that this change alters the guarantee of
the partition only slightly: We will obtain clusters bounded by
$\left(1+1/n^2\right)\Delta$, and in the estimate on the padding
probability the radii of the balls is changed by only a factor of
$(1\pm 1/n)$). We also do not process each scale, but rather work in
``event driven mode'': Vertices of $H$ are put in a non decreasing
order according to their labels in a queue. Each time we pop a new
vertex $u$, and partition the spaces at all the scales in the range
$[\Delta(u),n^2 \Delta(u)]$, for which we have not done so already.
In doing so we effectively skip ``irrelevant" scales. To estimate
the running time of this procedure note that the CKR decomposition
at scale $8^j$ takes time $O\left(m_i^2\right)$, where $m_i$ is the
number of vertices $u$ of $H$ satisfying~\eqref{eq:tree condition}
with $\Delta=8^i$. Note also that each vertex of $H$ participates in
at most $O(\log n)$ such CKR decompositions, so $\sum_i m_i=O(n\log
n)$. Hence the running time of the sampling procedure in
Lemma~\ref{lem:sample} is up to a constant factor $\sum_i
m_i^2=O\left(n^2\log n\right)$.

The Ramsey chain in Lemma~\ref{lem:sample} will be used in two
different ways in the ensuing constructions. For our approximate
distance oracle data structure we will just need that the
ultrametric $\rho_j$ is defined on $X_{j-1}$ (and not all of $X$).
Thus, by the above argument, and Lemma~\ref{lem:sample}, the
expected preprocessing time in this case is
$O\left(\mathbb{E}\sum_{j=1}^{s-1}|X_j|^2\log
|X_j|\right)=O\left(n^{2+1/k}\log n\right)$\, and the expected
storage space is $O\left(\mathbb{E}\sum_{j=1}^{s-1}|X_j|
\right)=O\left(n^{1+1/k}\right)$. For the purpose of our
approximate ranking data structure we will really need the metrics
$\rho_j$ to be defined on all of $X$. Thus in this case the
expected preprocessing time will be $O\left(n^2\log
n\cdot\mathbb{E}s\right)=O\left(kn^{2+1/k}\log n\right)$, and the
expected storage space is
$O\left(n\cdot\mathbb{E}s\right)=O\left(kn^{1+1/k}\right)$.

\paragraph{1) Approximate distance oracles.} Our improved approximate distance oracle is contained in
Theorem~\ref{thm:oracle}, which we now prove.

\begin{proof}[Proof of Theorem~\ref{thm:oracle}] We shall use the
notation in the statement of Lemma~\ref{lem:sample}. Let
$T_j=(V_j,E_j)$ and $\Delta_j:V_j\to (0,\infty)$  be the HST
representation of the ultrametric $\rho_j$ (which was actually
constructed explicitly in the proofs of Lemma~\ref{lem:ramseypart}
and Lemma~\ref{lem:sample}). The usefulness of the tree
representation stems from the fact that it very easy to handle
algorithmically. In particular there exists a simple scheme that
takes a tree and preprocesses it in linear time so that it is
possible to compute the least common ancestor of two given nodes
in constant time (see~\cite{HT84,BF00}). Hence, we can preprocess
any $1$-HST so that the distance between every two points can be
computed in $O(1)$ time.

For every point $x\in X$ let $i_x$ be the largest index for which
$x\in X_{i_x-1}$. Thus, in particular, $x\in Y_{i_x}$. We further
maintain for every $x\in X$ a vector (in the sense of
data-structures) $\mathbf{vec}_x$ of length $i_x$ (with $O(1)$
time direct access), such that for $i\in \{0,\ldots,i_x-1\}$,
$\mathbf{vec}_x[i]$ is a pointer to the leaf representing $x$ in
$T_i$. Now, given a query $x,y\in X$ assume without loss of
generality that $i_x\le i_y$. It follows that $x,y\in X_{i_x-1}$.
We locate the leaves $\hat{x}=\mathbf{vec}_x[i_x]$, and
$\hat{y}=\mathbf{vec}_y[i_x]$ in $T_{i_x}$, and then compute
$\Delta(\mathbf{lca}\left(\hat x,\hat y\right))$ to obtain an
$O(k)$ approximation to $d_X(x,y)$. Observe that the above data
structure only requires $\rho_j$ to be defined on $X_{j-1}$ (and
satisfying the conclusion of Lemma~\ref{lem:sample} for $x,y\in
X_{j-1}$).
% By the discussion following Lemma~\ref{prop:ramtopart},
The expected
preprocessing time is $O\left(n^{2+1/k} \log n\right)$.  The size
of the above data structure is $O\left(\sum_{j=0}^s |X_j|\right)$,
which is in expectation $O\left(n^{1+1/k}\right)$.
\end{proof}

\begin{remark}{Using the distributed labeling for the least common ancestor
operation on trees of Peleg~\cite{Peleg04}, the procedure described
in the proof of Theorem~\ref{thm:oracle} can be easily converted to
a {\em distance labeling} data structure (we refer
to~\cite[Section~3.5]{TZ05} for a description of this problem). We
shall not pursue this direction here, since while the resulting data
structure is non-trivial, it does not seem to improve over the known
distance labeling schema~\cite{TZ05}.}
\end{remark}

\begin{figure}[h]
\begin{center} \includegraphics{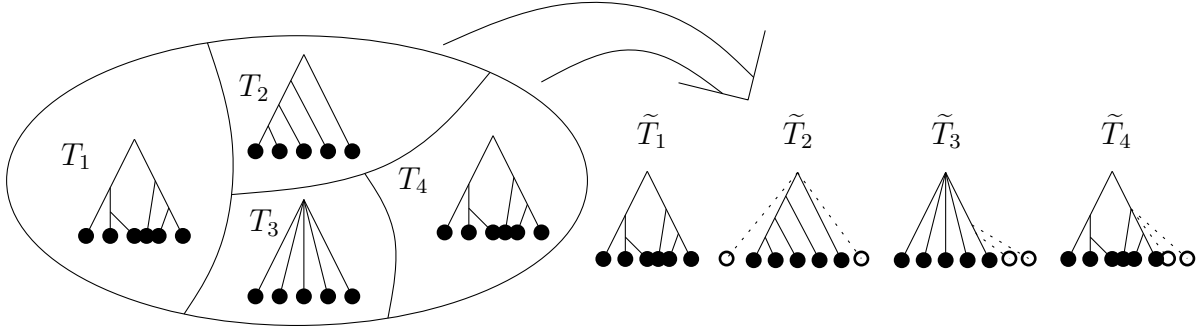} \end{center}
 \caption{\small{\em A schematic description of Ramsey
 chains and the way they are used to construct approximate
 distance oracles and approximate ranking data structures.} Ramsey chains are obtained by iteratively
 applying Theorem~\ref{thm:part-tree} and
Lemma~\ref{lem:ramseypart} to find a decreasing chain of subsets
$X=X_0\supsetneqq X_1\supsetneqq X_2\cdots \supsetneqq
X_s=\emptyset$ such that $X_j$ can be approximated by a  tree
metric $T_{j+1}$. The tree $T_{j+1}$ is, in a sense, a ``distance
estimator" for $X_j\setminus X_{j+1}$
--- it can be used to approximately evaluate the distance from a
point in $X_j\setminus X_{j+1}$ to any other point in $X_j$. These
trees form an array which is an approximate distance oracle. In
the case of approximate ranking we also need to extend the tree
$T_{j+1}$ to a tree on the entire space $X$ using
Lemma~\ref{lem:extend}. The nodes that were added to these trees
are illustrated by empty circles, and the dotted lines are their
connections to the original tree.
 } \label{fig:oracles}
\end{figure}

\paragraph{2) Approximate ranking.}
Before proceeding to our $\alpha$-approximate ranking data
structure (Theorem~\ref{thm:ranking}) we recall the setting of the
problem. Thinking of $X$ as a metric on $\{1,\ldots,n\}$, and
fixing $\alpha>1$, the goal here is to associate with every $x\in
X$ a permutation $\pi^{(x)}$ of $\{1,\ldots,n\}$ such that
$d_X\bigl(x,\pi^{(x)}(i)\bigr)\leq \alpha\cdot
d_X\bigl(x,\pi^{(x)}(j)\bigr)$ for every $1\leq i\leq j\leq n$.
This relaxation of the exact proximity ranking induced by the
metric $d_X$ allows us to gain storage efficiency, while enabling
fast access to this data. By fast access we mean that we can
preform the following tasks:
\begin{compactenum}
\item Given an element $x\in X$, and $i\in \{1,\ldots,n\}$, find $\pi^{(x)}(i)$ in $O(1)$ time.
\item Given an element $x\in X$ and $y\in X$, find number $i\in \{1,\ldots,n\}$, such that $\pi^{(x)}(i)=y$, in $O(1)$ time.
\end{compactenum}

\medskip

%Before passing to the proof of Theorem~\ref{thm:ranking}
We also require the following lemma.

\begin{lemma}\label{lem:data}
Let $T=(V,E)$ be a rooted tree with $n$ leaves. For $v\in V$, let
$\mathscr L_T(v)$ be the set of leaves in the subtree rooted at
$v$, and denote $\ell_T(v)=\left|\mathscr L_T(v)\right|$. Then
there exists a data structure, that we call
\textsf{Size-Ancestor}, which can be constructed in time $O(n)$,
so as to answer in time $O(1)$ the following query: Given $\ell\in
\mathbb N $ and a leaf $x\in V$, find an ancestor $u$ of $x$ such
that $\ell_T(u)<\ell\le \ell(\mathrm{parent}(u))$. Here we use the
convention
 $\ell(\mathrm{parent}(\mathrm{root}))=\infty$.
\end{lemma}

To the best of our knowledge, the data structure described in
Lemma~\ref{lem:data} has not been previously studied. We therefore
include a proof of Lemma~\ref{lem:data} in
Appendix~\ref{sec:size-ances}, and proceed at this point to conclude
the proof of Theorem~\ref{thm:ranking}.

\begin{proof}[Proof of Theorem~\ref{thm:ranking}] We shall use the
notation in the statement of Lemma~\ref{lem:sample}. Let
$T_j=(V_j,E_j)$ and $\Delta_j:V_j\to (0,\infty)$ be the HST
representation of the ultrametric $\rho_j$. We may assume without
loss of generality that each of these trees is binary and does not
contain a vertex which has only one child. Before presenting the
actual implementation of the data structure, let us explicitly
describe the permutation $\pi^{(x)}$ that the data structure will
use. For every internal vertex $v\in V_j$ assign arbitrarily the
value $0$ to one of its children, and the value $1$ to the other.
This induces a unique (lexicographical) order on the leaves of
$T_j$.
 Next, fix $x\in X$ and
$i_x$ such that $x\in Y_{i_x}$. The permutation $\pi^{(x)}$ is
defined as follows. Starting from the leaf $x$ in $T_{i_x}$, we scan
the path from $x$ to the root of $T_{i_x}$. On the way, when we
reach a vertex $u$ from its child $v$, let $w$ denote the sibling of
$v$, i.e. the other child of $u$. We next output all the leafs which
are descendants of $w$ according to the total order described above.
Continuing in this manner until we reach the root of $T_{i_x}$ we
obtain a permutation $\pi^{(x)}$ of $X$.

We claim that the permutation $\pi^{(x)}$ constructed above is an
$O(k)$-approximation to the proximity ranking induced by $x$.
Indeed, fix $y,z\in X$ such that $Ck \cdot d_X(x,y) <
d_X(x,z)$, where $C$ is a large enough absolute constant. We claim
that $z$ will appear after $y$ in the order induced by
$\pi^{(x)}$. This is true since the distances from $x$ are
preserved up to a factor of $O(k)$ in the ultrametric $T_{i_x}$.
Thus for large enough $C$ we are guaranteed that
$d_{T_{i_x}}(x,y)< d_{T_{i_x}}(x,z)$, and therefore
$\lca_{T_{i_x}}(x,z)$ is a proper ancestor of
$\lca_{T_{i_x}}(x,y)$. Hence in the order just describe above, $y$
will be scanned before $z$.

We now turn to the description of the actual data structure, which
is an enhancement of the data structure constructed in the proof
of Theorem~\ref{thm:oracle}. As in the proof of
Theorem~\ref{thm:oracle} our data structure will consist of a
``vector of the trees $T_j$", where we maintain for each $x\in X$
a pointer to the leaf representing $x$ in each $T_j$. The
remaining description of our data structure will deal with each
tree $T_j$ separately. First of all, with each vertex $v\in T_j$
we also store the number of leaves which are the descendants of
$v$, i.e. $|\mathscr L_{T_j}(v)|$ (note that all these numbers can
be computed in  $O(n)$ time using, say, depth-first search). With
each leaf of $T_j$ we also store its index in the order described
above. There is a reverse indexing by a vector for each tree $T_j$
that allows, given an index, to find the corresponding leaf of
$T_j$ in $O(1)$ time. Each internal vertex contains a pointer to
its leftmost (smallest) and rightmost (largest) descendant leaves.
This data structure can be clearly constructed in $O(n)$ time
using, e.g., depth-first transversal of the tree. We now give
details on how to answer the required queries using the
``ammunition" we have listed above.

\begin{enumerate}
\item Using Lemma~\ref{lem:data}, find an ancestor $v$ of $x$ such that
$\ell_{T_j}(v)< i \le \ell_{T_j}(\text{parent}(v))$ in $O(1)$
time. Let $u=\text{parent}(v)$ (note that $v$ can not be the root).
Let $w$ be the sibling of $v$
(i.e. the other child of $u$). Next we pick the leaf numbered
$\left(i-\ell_{T_j}(v)\right)+\text{left}(w)-1$, where
$\text{left}(w)$ is the index to the leftmost descendant of $w$.

\item
Find $u=\lca(x,y)$ (in $O(1)$ time, using~\cite{HT84,BF00}). Let
$v$ and $w$ be the children of $u$, which are ancestors of $x$ and
$y$, respectively. Return $\ell_{T_j}(v)+
\text{ind}(y)-\text{left}(w)$, where $\text{ind}(y)$ is the index
of $y$ in the total order of the leaves of the tree. \remove{
\item We will describe how to go upward in $\pi^{(x)}$. Going downward is suppose to be similar. Starting from $y$; let $v,w$ as in
the previous items. Output the elements with the indices $\text{ind}(y)$ up to $\text{ind}(\text{right}(w))$
(if we haven't reached yet the quota of $t$). After that we $v\leftarrow u$, $u\leftarrow \text{parent}(v)$,
$w\leftarrow \text{sibling}(v)$, and continue this way until we reach the quota of $v$.
}
\end{enumerate}

This concludes the construction of our approximate ranking data
structure.
%As discussed after Lemma~\ref{lem:ramtopart},
Because
we need to have the ultrametric $\rho_j$ defined on all of $X$,
the preprocessing time is $O\left(kn^{2+1/k}\log n\right)$ and the
storage size is $O\left(kn^{1+1/k}\right)$, as required.
\end{proof}

\begin{remark} {Our approximate ranking data structure
 can also be used in a nearest neighbor heuristic called  ``Orchard
Algorithm" \cite{Orc91} (see also~\cite[Sec.~3.2]{Clarkson05}). In
this algorithm the vanilla heuristic can be used to obtain the exact
proximity ranking, and requires storage $\Omega\left(n^2\right)$.
Using approximate ranking the storage requirement can be
significantly improved, though the query performance is somewhat
weaker due to the inaccuracy of the ranking lists.}
\end{remark}

\paragraph{3) Computing the Lipschitz constant.} Here we describe a data
structure for computing the Lipschitz constant of a function $f:
X\to Y$, where $(Y,d_Y)$ is an arbitrary metric space. When
$(X,d_X)$ is a {\em doubling metric space} (see~\cite{Hei01}), this
problem was studied in~\cite{HM05}. In what follows we shall always
assume that $f$ is given in {\em oracle form}, i.e. it is encoded in
such a way that we can compute its value on a given point in
constant time.

\begin{lemma}\label{lem:Lip-um} There is an algorithm that, given
an $n$-point ultrametric $(U,d_U)$ defined by the HST $T=(V,E)$
(in particular $U$ is the set of leaves of $T$), an arbitrary
metric space $(Y,d_Y)$, and a mapping $f:U \to Y$, returns in time
$O(n)$ a number $A\geq 0$ satisfying $\|f\|_{\mathrm{Lip}} \geq A
\geq \frac{1}{16}\cdot\|f\|_{\mathrm{Lip}}$.
\end{lemma}
\begin{proof}
We assume that $T$ is 4-HST. As remarked in the beginning of
Section~\ref{section:use}, this can be achieved by distorting the
distances in $U$ by a factor of at most $4$, in $O(n)$ time. We
also assume that the tree $T$ stores for every vertex $v\in V$ an
arbitrary leaf $x_v\in U$ which is a descendant of $v$ (this can
be easily computed in $O(n)$ time). For a vertex $u\in V$ we
denote by $\Delta(u)$ its label (i.e. $\forall\ x,y\in U,\
d_U(x,y)=\Delta(\lca(x,y))$).

The algorithm is as follows:

\begin{center} \fbox{
\begin{minipage}{0.9\columnwidth}
\begin{tabbing}
\ \ \ \= \ \ \ \= \ \ \ \= \kill\\
\quad\quad\quad\quad\quad \underline{\bf{Lip-UM}$(T,f)$} \+\\
$A\leftarrow 0$ \\
For every vertex $u\in T$ \textbf{do} \+\\
Let $v_1,\ldots v_r$ be the children of $u$.\\
$A\leftarrow \max \left\{A, \max_{2\leq i\leq r}
\frac{d_Y\left(f(x_{v_1}),f(x_{v_i})\right)} {\Delta(u)} \right \}$ \-\\
Output $A$.
\end{tabbing}
\end{minipage}
}
\end{center}

\bigskip

 Clearly the algorithm runs in linear time (the total
number of vertices in the tree is $O(n)$ and each vertex is
visited at most twice). Furthermore, by construction the algorithm
outputs $A\leq \|f\|_{\text{Lip}}$. It remains to prove a lower
bound on $A$. Let $x_1,x_2\in U$ be such that
$\|f\|_{\text{Lip}}=\frac{d_Y(f(x_1),f(x_2))}{d_U(x_1,x_2)}$, and
denote $u=\lca(x,y)$. Let $w_1,w_2$ be the children of $u$ such
that $x_1\in \mathscr L_T(w_1)$, and $x_2\in \mathscr L_T(w_2)$.
Let $v_1$ be the ``first child" of $u$ as ordered by the algorithm
Lip-UM (notice that this vertex has special role). Then
\begin{eqnarray*}
A &\geq& \max\left \{
\frac{d_Y(f(x_{w_1}),f(x_{v_1}))}{\Delta(u)},
\frac{d_Y(f(x_{w_2}),f(x_{v_1}))}{\Delta(u)} \right \} \\
& \geq& \frac{1}{2}\cdot
\frac{d_Y(f(x_{w_1}),f(x_{w_2}))}{\Delta(u)}\\
& \geq& \frac{1}{2}\cdot \frac{d_Y(f(x_1),f(x_2))-\diam(f(\mathscr
L_T(w_1)))-\diam(f(\mathscr L_T(w_2)))}{\Delta(u)}.
\end{eqnarray*}
If $\max\left\{\diam(f(\mathscr L_T(w_1))),\diam(f(\mathscr
L_T(w_2))) \right\} \leq \frac{1}{4} d_Y(f(x_1),f(x_2))$, then we
conclude that
\[ A \geq \frac{1}{4}\cdot \frac{d_Y(f(x_1),f(x_2))}{\Delta(u)},   \]
as needed. Otherwise, assuming  that $\diam(f(\mathscr L_T(w_1)))>
\frac14 \cdot d_Y(f(x_1),f(x_2))$, there exist $z,z'\in \mathscr
L_T(w_1)$ such that
\[ \frac{d_Y(f(z),f(z'))}{d_U(z,z')} > \frac{\frac14 \cdot d_Y(f(x_1),f(x_2))}
{\Delta(u)/4}= \|f\|_{\text{Lip}} ,\] which is a contradiction.
\end{proof}

\begin{theorem} Given $k\ge 1$, any $n$-point metric space $(X,d_X)$ can be preprocessed in
time $O\left(n^{2+1/k}\log n\right)$, yielding a data structure
requiring storage $O\left(n^{1+1/k}\right)$ which can answer in
$O\left(n^{1+1/k}\right)$ time the following query: Given a metric
space $(Y,d_Y)$ and a mapping $f:X \to Y$, compute a value $A\ge
0$, such that $\|f\|_{\mathrm{Lip}} \geq A \geq
\|f\|_{\mathrm{Lip}} / O(k)$.
\end{theorem}
\begin{proof}
The preprocessing is simply computing the trees $\{T_j\}_{j=1}^s$
as in the proof of Theorem~\ref{thm:oracle}. Denote the resulting
ultrametrics by $(U_1,\rho_1),\ldots, (U_s,\rho_s)$. Given $f:X\to
Y$, represent it as $g_i:U_i\to Y$ (as a mapping $g_i$ is the same
mapping as the restriction of $f$ to $U_i$). Use
Lemma~\ref{lem:Lip-um} to compute an estimate $A_i$ of
$\|g_i\|_{\mathrm{Lip}}$, and return $A\coloneqq \max_i A_i$.
Since all the distances in $U_i$ dominate the distances in $X$,
$\|f\|_{\mathrm{Lip}} \geq \|g_i\|_{\mathrm{Lip}} \geq A_i$, so
$\|f\|_{\mathrm{Lip}}\geq A$. On the other hand, let $x,y\in X$ be
such that $\|f\|_{\mathrm{Lip}} = \frac{d_Y(f(x),f(y))}
{d_X(x,y)}$. By Lemma~\ref{lem:sample}, there exists $i\in
\{1,\ldots,s\}$ such that $d_{U_i}(x,y)\leq O(k)\cdot d_X(x,y)$,
and hence $\|g_i\|_{\mathrm{Lip}} \geq \|f\|_{\mathrm{Lip}}/O(k)$,
And so $A \geq \frac{1}{16}\cdot\|g_i\|_{\mathrm{Lip}} \geq
\|f\|_{\mathrm{Lip}}/O(k)$, as required. Since we once more only
need that the ultrametric $\rho_{j}$ is defined on $X_{j-1}$ and
not on all of $X$, the preprocessing time and storage space are
the same as in Theorem~\ref{thm:oracle}. By Lemma~\ref{lem:sample}
the query time is
$O\left(\sum_{j=1}^{s-1}|X_j|\right)=O\left(n^{1+1/k}\right)$ (we
have a $O(|X_{j}|)$ time computation of the Lipschitz constant on
each $X_j$).
\end{proof}

\section{Concluding Remarks}

An $s$-\emph{well separated pair decomposition} (WSPD) of an
$n$-point  metric space $(X,d_X)$ is a collection of pair of
subsets $\{(A_i,B_i)\}_{i=1}^M$, $A_i,B_i\subset X$, such that
\begin{enumerate}
\item $\forall x,y\in X$ if  $x\neq y$ then $(x,y) \in \bigcup_{i=1}^M (A_i \times B_i)$.
\item For all $i\neq j$, $(A_i\times B_i) \cap (A_j\times B_j) = \emptyset$.
\item For all $i\in \{1,\ldots,M\}$, $ d_X(A_i,B_i)\ge s\cdot \max\{\diam(A_i),\diam(B_i)\}$.
\end{enumerate}
The notion of $s$-WSPD was first defined for Euclidean spaces in
an influential paper of Callahan and Kosaraju~\cite{CK95}, where
it was shown that for $n$-point subsets of a fixed dimensional
Euclidean space there exists such a collection of size $O(n)$ that
can be constructed in $O(n \log n)$ time. Subsequently, this
concept has been used in many geometric algorithms
(e.g.~\cite{Vai89,Call95}), and is today considered to be a basic
tool in computational geometry. Recently the definition and the
efficient construction of WSPD were generalized to the more
abstract setting of doubling metrics~\cite{Tal04,HM05}. These
papers have further demonstrated the usefulness of this tool (see
also~\cite{FK05} for a mathematical application).

 It would be clearly desirable to have a
notion similar to WSPD in general metrics. However, as formulated
above, no non-trivial WSPD  is possible in ``high dimensional"
spaces, since any $2$-WSPD of an $n$-point equilateral space must
be of size $\Omega(n^2)$. The present paper suggests that Ramsey
partitions might be a partial replacement of this notion which
works for arbitrary metric spaces. Indeed, among the applications
of WSPD in fixed dimensional metrics are approximate ranking
(though this application does not seem to have appeared in print
--- it was pointed out to us by Sariel Har-Peled),  approximate
distance oracles~\cite{Gud02,HM05}, spanners~\cite{Tal04,HM05},
and computation of the Lipschitz constant~\cite{HM05}. These
applications have been obtained for general metrics using Ramsey
partitions in the present paper (spanners were not discussed here
since our approach does not seem to beat previously known
constructions). We believe that this direction deserves further
scrutiny, as there are more applications of WSPD which might be
transferable to general metrics using Ramsey partitions. With is
in mind it is worthwhile to note here that our procedure for
constructing stochastic Ramsey chains, as presented in
Section~\ref{section:use}, takes roughly $n^{2+1/k}$ time (up to
logarithmic terms). For applications it would be desirable to
improve this construction time to $O(n^2)$. The construction time
of ceratin proximity data structures is a well studied topic in
the computer science literature --- see for
example~\cite{Z01,LTZ05}.

\paragraph{Acknowledgments.} Part of this work
was carried out while Manor Mendel was visiting Microsoft
Research. We are grateful to Sariel Har-Peled for letting us use
here his insights on the approximate ranking problem. We also
thank Yair Bartal for helpful discussions.

\appendix
\section*{Appendices}

\section{The Size-Ancestor data structure} \label{sec:size-ances}

In this  appendix we prove Lemma~\ref{lem:data}. Without loss of
generality we assume that the tree $T$  does not contain vertices
with only one child. Indeed, such vertices will never be returned as
an answer for a query, and thus can be eliminated in $O(n)$ time in
a preprocessing step.

Our data structure
is composed in a modular way of two different data structures, the
first of which is described in the following lemma, while the second
is discussed  in the proof of Lemma~\ref{lem:data} that will follow.

\begin{lemma}\label{lem:size_ances_s}
Fix $m\in\mathbb{N}$, and let $T$ be as in Lemma~\ref{lem:data}.
Then there exists a data structure which can be preprocessed in time
$O\left(n +\frac{n \log n}{m}\right)$, and answers in time $O(1)$
the following query: Given $\ell\in \mathbb N $ and a leaf $x\in V$,
find an ancestor $u$ of $x$ such that $\ell_T(u)<\ell m\le
\ell(\mathrm{parent}(u))$. Here we use the convention
$\ell(\mathrm{parent}(\mathrm{root}))=\infty$.
\end{lemma}
\begin{proof} Denote by $X$ the set of leaves of $T$.
For every internal vertex $v\in V$, order its children
non-increasingly according to the number of leaves in the subtrees
rooted at them. Such a choice of labels induces a unique total
order on $X$ (the lexicographic order). Denote this order by
$\preccurlyeq$ and let $f: \{1,\ldots,n\}\to X$ be the unique
increasing map in the total order $\preccurlyeq$. For every $v\in
V$, $f^{-1}\left(\mathscr L_T(v)\right)$ is an interval of
integers. Moreover, the set of intervals $\bigl\{f^{-1}(\mathscr
L_T(v)):\ v\in V\bigr\}$ forms a laminar set, i.e. for every pair
of intervals in this set either one is contained in the other, or
they are disjoint. For every $v\in V$ write $f^{-1}\left(\mathscr
L_T(v)\right)=I_v=[A_v,B_v]$, where $A_v,B_v\in \mathbb N$ and
$A_v\le B_v$. For $i\in \{1,\ldots,\lfloor n/m\rfloor\}$ and $j\in
\{1,\ldots,\lceil n/(im)\rceil\}$ let $F_i(j)$ be the set of
vertices $v\in V$ such that $|I_v|\ge im$, $I_v\cap
[(j-1)im+1,jim]\neq \emptyset $, and there is no descendant of $v$
satisfying these two conditions. Since at most two disjoint
intervals of length at least $im$ can intersect a given interval
of length $im$, we see that for all $i,j$, $|F_i(j)|\le 2$.

\begin{claim}\label{claim:step} Let $x\in X$ be a leaf of $T$, and $\ell\in \mathbb
N$. Let $u\in V$ be the least ancestor of $x$ for which
$\ell_T(u)\ge \ell m$. Then
$$
u\in \left\{\lca(x,v):\ v\in F_\ell\left(\left\lceil
\frac{f(x)}{\ell m}\right\rceil\right)\right\}.
$$
\end{claim}
\begin{proof} If $u\in F_\ell\left(\left\lceil
\frac{f(x)}{\ell m}\right\rceil\right)$ then since $u=\lca(x,u)$
there is nothing to prove. If on the other hand $u\notin
F_\ell\left(\left\lceil \frac{f(x)}{\ell m}\right\rceil\right)$
then since we are assuming that $\ell_T(u)\ge \ell m$, and
$I_u\cap \left[\left(\left\lceil \frac{f(x)}{\ell m
}\right\rceil-1\right)\ell m+1,\left\lceil \frac{f(x)}{\ell m}
\right\rceil\ell m\right]\neq \emptyset $ (because $f(x)\in I_u$),
it follows that $u$ has a descendant $v$ in
$F_\ell\left(\left\lceil \frac{f(x)}{\ell m}\right\rceil\right)$.
Thus $u=\lca(x,v)$, by the fact that {\em any} ancestor $w$ of $v$
satisfies $\ell_T(w)\ge \ell_T(v)\ge \ell m$, and the minimality
of $u$.
\end{proof}

% We can now present the data-structure.

The preprocessing of the data structure begins with
ordering the children of vertices
non-increasingly according to the number of leaves in their
subtrees. The following algorithm achieves it in linear time.

\begin{comment}
To do this we first compute
$\{\ell_T(u)\}_{u\in V}$ using depth first search.
We then sort \emph{all the vertices} according to
$\ell_T(\cdot)$ in $O(n)$ time using bucket sort
(see~\cite[Ch.~9]{CLR90}
 --- we are using here the fact that the $O(n)$
numbers $\{\ell_T(u)\}_{u\in V}$ are in the range $\{0,\ldots,n\}$).
Next, we initialize for each vertex $u$ a new empty list called
$\mathsf{ChildrenSortedList}_u$. And then scan our list of all
vertices sorted (decreasingly) according to $\ell_T(\cdot)$, and
when reaching a vertex $v$ in the list, we add $v$ to the end of
$\mathsf{ChildrenSortedList}_{\mathrm{parent}(v)}$. In the end of
this process, for each vertex $u$ the list
$\mathsf{ChildrenSortedList}_u$ is the list of its children, sorted
decreasingly according to $\ell_T(\cdot)$, as needed. Computing $f$,
and the intervals $\{I_u\}_{u\in V}$ is now done by a depth-first
scan of the tree that respects the order of the children.
\end{comment}

\begin{center} \fbox{
\begin{minipage}{\columnwidth}
\begin{tabbing}
\ \ \ \ \=  \ \ \ \ \= \ \ \ \ \= \ \ \ \ \=\kill \\
 \quad\quad\quad\quad\quad\quad\quad\quad\quad\quad\quad\underline{\bf{SORT-CHILDREN}$(u)$} \+\\
Compute $\{\ell_T(u)\}_{u\in V}$ using depth first search.\\
Sort $V$ non-increasingly according to $\ell_T(\cdot)$ (use bucket sort- see~\cite[Ch.~9]{CLR90}).\\
Let $(v_i)_i$ be the set $V$ sorted as above.\\
Initialize $\forall u\in V$, the list
$\mathsf{ChildrenSortedList}_{u}= \emptyset$.\\
For $i=1$ to $|V|$ \textbf{do} \\
\> Add $v_i$ to the end of $\mathsf{ChildrenSortedList}_{\mathrm{parent}(v_i)}$.
\end{tabbing}
\end{minipage}}
\end{center}

Computing $f$, and the intervals $\{I_u\}_{u\in V}$ is now done by a
depth first search  of $T$ that respects the above order of the
children. We next compute $\left\{F_i(j):\ i\in \{1,\ldots,\lfloor
n/m\rfloor\},\ j\in \{1,\ldots,\lceil n/(im)\rceil\right\}$ using
the following algorithm:

\begin{center} \fbox{
\begin{minipage}{\columnwidth}
\begin{tabbing}
\ \ \ \ \=  \ \ \ \ \= \ \ \ \ \= \ \ \ \ \=\kill \\
 \quad\quad\quad\quad\quad\quad\quad\quad\underline{\bf{SUBTREE-COUNT}$(u)$} \+\\
Let $v_1, \ldots, v_r$ be the children of $u$
with $|I_{v_1}|\geq |I_{v_2}| \geq \cdots \geq |I_{v_r}|$.\\
For $i\leftarrow \left\lfloor |I_{u}|/m\right\rfloor$ down to $\left\lfloor|I_{v_{1}}|/m\right\rfloor +1$ \textbf{do} \+ \\
For $j \leftarrow \lfloor A_{u}/(i m) \rfloor$ to
 $\lceil B_u/(im) \rceil$ \textbf{do} \+\\
Add $u$ to $F_i(j)$ \-\-\\
For $h \leftarrow 1$ to $r-1$ \textbf{do} \+\\
For $i\leftarrow \left\lfloor |I_{v_h}|/m\right\rfloor$ down to $\left\lfloor|I_{v_{h+1}}|/m\right\rfloor +1$ \textbf{do} \+ \\
For $j \leftarrow \lceil  B_{v_h}/(i m) \rceil+1$ to
 $\lceil B_u/(im) \rceil$ \textbf{do} \+\\
Add $u$ to $F_i(j)$ \-\-\-\\
For $h \leftarrow 1$ to $r$ \textbf{do} call SUBTREE-COUNT$(v_h)$.
\end{tabbing}
\end{minipage}
}
\end{center}
Here is an informal explanation of the correctness of this
algorithm. The only relevant sets $F_i(\cdot)$ which will contain the
vertex $u\in V$ are those in the range $i\in[\lfloor|I_{v_r}| /m
\rfloor+1, \lfloor |I_{u}| /m \rfloor ]$. Above this range $I_u$
does not meet the size constraint, and below this range any $F_i(j)$
which intersects $I_u$ must also intersect one of the children of
$u$, which also satisfies the size constraint, in which case one of
the descendants of $u$ will be in $F_i(j)$. In the aforementioned
range, we add $u$ to $F_i(j)$ only for $j$ such that the interval
$[(j-1)im+1,jim]$ does not intersect one of the children of $u$ in a
set of size larger than $im$. Here we use the fact that the
intervals of the children are sorted in non-increasing order
according to their size. Regarding running time, this reasoning
implies that each vertex of $T$, and each entry in $F_i(j)$, is
accessed by this algorithm only a constant number of times, and each
access involves only constant number of computation steps. So the
running time is
\[O\Biggl(n+\sum_{i=1}^{\lfloor n/m\rfloor}\sum_{j=1}^{\lceil
n/(im)\rceil}|F_i(j)|\Biggr)=O\left(n+\frac{n\log n}{m}\right).\]

We conclude with the query procedure. Given a query $x\in X$ and
$\ell\in\mathbb{N}$, access $F_\ell\left(\left\lceil
\tfrac{f(x)}{\ell} \right\rceil\right)$ in $O(1)$ time. Next, for
each $v\in F_\ell\left(\left\lceil \tfrac{f(x)}{\ell}
\right\rceil\right)$, check whether $\lca(x,v)$ is the required
vertex (we are thus using here also the data structure for
computing the $\lca$ of~\cite{HT84,BF00}. Observe also that since
$|F_i(j)|\leq 2$, we only have a constant number of checks to do).
By Claim~\ref{claim:step} this will yield the required result.
\end{proof}

By setting $m=1$ in Lemma~\ref{lem:size_ances_s}, we obtain a data
structure for the \textsf{Size-Ancestor} problem with $O(1)$ query
time, but $O(n \log n)$ preprocessing time. To improve
upon this, we set $m=\Theta(\log n)$ in
Lemma~\ref{lem:size_ances_s}, and deal with the resulting gaps by
enumerating all the possible ways in which the remaining $m-1$
leaves can be added to the tree. Exact details are given below.

\begin{proof}[Proof of Lemma~\ref{lem:data}]
Fix $m=\lfloor (\log n)/4 \rfloor$. Each subset $A\subseteq
\{0,\ldots,m-1\}$ is represented as a number $\#A\in
\{0,\ldots,2^m-1\}$ by
\( \#A = \sum_{i\in A} 2^i. \)
We next construct in memory a vector $\mathbf{enum}$ of size
$2^{m}$, where $\mathbf{enum}[\#A]$ is a vector of size $m$, with
integer index in the range $\{1,\ldots,m\}$,
 such that $\mathbf{enum}[\#A][i]= |A\cap
\{0,\ldots,i-1\}|$. Clearly $\mathbf{enum}$ can be constructed in
$O(2^{m} m)=o(n)$ time.

For each vertex $u$ we compute and store:
\begin{compactitem}
\item $\mathrm{depth}(u)$ which is the edge's distance from the root to $u$.
\item $\ell_T(u)$, the number of of leaves in the subtree rooted at $u$.
\item The number $\#A_u$, where
\[ A_u=\Bigl\{k\in\{0,\ldots,m-1\}:\ u \text{ has an ancestor with exactly }
\ell_T(u)+k \text{ descendant leaves} \Bigr\} .\]
\end{compactitem}
We also apply the level ancestor data-structure, that after $O(n)$
preprocessing time, answers in constant time queries of the form:
Given a vertex $u$ and an integer $d$, find an ancestor of $u$ at
depth $d$ (if it exists) (such a data structure is constructed
in~\cite{BF04b}). Lastly, we use the data structure from
Lemma~\ref{lem:size_ances_s}

With all this machinary in place, a query for the least ancestor of
a leaf $x$ having at least $\ell$ leaves is answered in constant
time as follows. First compute $q=\lfloor \ell/m\rfloor$. Apply a
query to the data structure of Lemma~\ref{lem:size_ances_s}, with
$x$ and $q$, and obtain $u$, the least ancestor of $x$ such that
$\ell_T(u)\geq qm$. If $\ell_T(u)\geq \ell$ then $u$ is the least
ancestor with $\ell$ leaves, so the data-structure returns $u$.
Otherwise, $\ell_T(u)<\ell$, and let
$a=\mathbf{enum}[\#A_u][\ell-\ell_T(u)]$. Note that
$\mathrm{depth}(u)-a$ is the depth of the least ancestor of $u$
having at least $\ell$ leaves, thus the query uses the level
ancestor data-structure to return this ancestor. Clearly the whole
query takes a constant time.

It remains to argue that the data structure can be preprocessed in
linear time. We already argued about most  parts of the data
structure, and $\ell_T(u)$ and $\mathrm{depth}(u)$ are easy to
compute in linear time. Thus we are left with computing $\#A_u$ for
each vertex $u$. This is done using a top-down scan of the tree
(e.g., depth first search). The root is assigned with 1.  Each
non-root vertex $u$, whose parent is $v$, is assigned
\[ \#A_u \leftarrow \begin{cases}
1 & \text { if } \ell_T(v)\geq \ell_T(u)+m \\
\#A_v \cdot 2^{\ell_T(v)-\ell_T(u)} +1 \pmod{2^m} & \text{
otherwise.}
\end{cases}
\]
It is clear that this indeed computes $\#A_u$.
The relevant exponents are computed in advance and stored in a lookup table.
% Hence, each vertex takes constant time.
\end{proof}

\begin{remark}
This data structure can be modified in a straightforward way to
answer queries to the least ancestor of a given size (in terms of
the number of vertices in its subtree). It is also easy to extend it
to queries which are non-leaf vertices.
\end{remark}

\section{The metric Ramsey theorem implies the existence of Ramsey
partitions} \label{app:converse}

In this appendix we complete the discussion in
Section~\ref{section:part} by showing that the metric Ramsey theorem
implies the existence of good Ramsey partitions. The results here
are not otherwise used in this paper.

\begin{prop}\label{prop:ramtopart} Fix $\alpha\ge 1$ and $\psi\in
(0,1)$, and assume that every $n$-point metric space has a subset
of size $n^\psi$ which is $\alpha$-equivalent to an ultrametric.
Then every $n$-point metric space $(X,d_X)$ admits a distribution
over partition trees $\left\{\mathscr R_k\right\}_{k=0}^\infty$
such that for every $x\in X$,
$$
\Pr\left[\forall\ k\in \mathbb N,\
B_X\left(x,\frac{1}{96\alpha}\cdot 8^{-k}\diam(X)\right)\subseteq
\mathscr R_k(x)\right]\ge \frac{1-\psi}{n^{1-\psi}}.
$$
\end{prop}

\begin{proof} Let $(X,d_X)$ be an $n$-point metric space.
The argument starts out similarly to the proof of
Lemma~\ref{lem:sample}. Using the assumptions
and Lemma~\ref{lem:extend} iteratively,
we find a decreasing chain of subsets $X=X_0\supsetneqq
X_1\supsetneqq X_2\cdots \supsetneqq X_s=\emptyset$ and
ultrametrics $\rho_1,\ldots,\rho_s$ on $X$, such that if we denote
$Y_j=X_{j-1}\setminus X_j$ then $|Y_{j}|\ge |X_{j-1}|^\psi$, for
$x,y\in X$, $\rho_j(x,y)\ge d_X(x,y)$, and for $x\in X$, $y\in
Y_j$ we have $\rho_j(x,y)\le 6\alpha  d_X(x,y)$. As in the proof
of Lemma~\ref{lem:sample}, it follows by induction that $s\le
\frac{1}{1-\psi}\cdot n^{1-\psi}$.

By~\cite[Lemma 3.5]{BLMN05} we may assume that the ultrametric
$\rho_j$ can be represented by an exact $2$-HST $T_j=(V_j,E_j)$,
with vertex labels $\Delta_{T_j}$, at the expense of replacing the
factor $6$ above by $12$. Let $\Delta_j$ be the label of the root
of $T_j$, and denote for $k\in \mathbb N$, $\Lambda_j^k=\{v\in
V_j:\ \Delta_{T_j}(v)=2^{-k}\Delta_j\}$. For every $v\in V_j$ let
$\mathscr L_j(v)$ be the leaves of $T_j$ which are descendants of
$v$. Thus $\mathscr P^k_j\coloneqq \left\{\mathscr L_j(v):\ v\in
\Lambda_j^k\right\}$ is a $ 2^{-k}\Delta_j$ bounded partition of
$X$ (boundedness is in the metric $d_X$). Fix $x\in Y_j$, $k\in
\mathbb{N}$ and let $v$ be the unique ancestor of $X$ in
$\Lambda_j^k$. If $z\in X$ is such that $d_X(x,z)\le
\frac{1}{12\alpha}\cdot 2^{-k}{\Delta_j}$ then
$\Delta_{T_j}\left(\lca_{T_j}(x,z)\right)=\rho_j(x,z)\le
2^{-k}\Delta_j$. It follows that $z$ is a descendant of $v$, so
that $z\in \mathscr P_j^k(x)=\mathscr L_j(v)$. Thus $\mathscr
P_j^k(x)\supseteq B_X\left(x,\frac{1}{12\alpha}\cdot
2^{-k}{\Delta_j}\right)$.

Passing to powers of $8$ (i.e. choosing for each $k$ the integer
$\ell$ such that $8^{-\ell-1}\diam(x)<2^{-k}\Delta_j\le
8^{-\ell}\diam(X)$ and indexing the above partitions using $\ell$
instead of $k$), we have thus shown that for every $j\in
\{1,\ldots,s\}$ there is a partition tree $\left\{\mathscr
R_k^j\right\}_{k=0}^\infty$ such that for every $x\in Y_j$ we have
for all $k$,
$$
B_X\left(x,\frac{1}{96\alpha}\cdot 8^{-k}\diam(X)\right)\subseteq
\mathscr R_k^j(x).
$$
Since the sets $Y_1,\ldots,Y_s$ cover $X$, and $s\le
\frac{n^{1-\psi}}{1-\psi}$, the required distribution over
partition trees can be obtained by choosing one of the partition
trees $\left\{\mathscr
R_k^1\right\}_{k=0}^\infty,\ldots,\left\{\mathscr
R_k^s\right\}_{k=0}^\infty$ uniformly at random.
\end{proof}

\begin{remark}{ Motivated by the re-weighting argument in~\cite{CGR05},
it is possible to improve the lower bound in
Proposition~\ref{prop:ramtopart} for $\psi$ in a certain range. We
shall now sketch this argument. It should be remarked, however, that
there are several variants of this re-weighting procedure (like
re-weighting again at each step), so it might be possible to
slightly improve upon Proposition~\ref{prop:ramtopart} for a larger
range of $\psi$. We did not attempt to optimize this argument here.

Fix $\eta\in (0,1)$ to be determined in the ensuing argument, and
let $(X,d_X)$ be an $n$-point metric space. Duplicate each point
in $X$ $n^\eta$ times, obtaining a (semi-) metric space $X'$ with
$n^{1+\eta}$ points (this can be made into a metric by applying an
arbitrarily small perturbation). We shall define inductively a
decreasing chain of subsets $X'=X_0'\supsetneqq X_1'\supsetneqq
X_2'\supsetneqq \cdots$ as follows. For $x\in X$, let $h_i(x)$ be
the number of copies of $x$ in $X_i'$ (thus $h_0(x)=n^{\eta}$).
Having defined $X_i'$, let $Y_{i+1}\subseteq X_i'$ be a subset
which is $\alpha$-equivalent to an ultrametric and $|Y_{i+1}|\ge
|X_i'|^\psi$. We then define $X_{i+1}'$ via
\[ h_{i+1}(x)=\begin{cases} \lfloor h_i(x)/2 \rfloor &  \mathrm{there\  exists\  a\  copy\  of\ }  x\ \mathrm{in\ } Y_{i+1} \\
h_i(x) & \text{otherwise} . \end{cases}  \]

Continue this procedure until we arrive at the empty set. Observe
that $$|X_{i+1}'|\le |X'_i|-\frac12|X_i'|^\psi\le
|X_i'|\left(1-\frac{1}{2n^{(1+\eta)(1-\psi)}}\right).$$ Thus
$|X_i'|\le n^{1+\eta}\cdot
\left(1-\frac{1}{2n^{(1+\eta)(1-\psi)}}\right)^{i-1}$. It follows
that this procedure terminates after
$O\left(n^{(1+\eta)(1-\psi)}\log n\right)$ steps, and by
construction each point of $X$ appears in $\Theta\left( \eta \log
n\right)$ of the subsets $Y_i$. As in the proof of
Proposition~\ref{prop:ramtopart}, by selecting each of the $Y_i$
uniformly at random we get a distribution over partition trees
$\left\{\mathscr R_k\right\}_{k=0}^\infty$ such that for every $x\in
X$,
$$
\Pr\left[\forall\ k\in \mathbb N,\
B_X\left(x,\frac{1}{96\alpha}\cdot 8^{-k}\diam(X)\right)\subseteq
\mathscr R_k(x)\right]\ge
\Omega\left(\frac{\eta}{n^{(1+\eta)(1-\psi)}}\right).
$$
Optimizing over $\eta\in (0,1)$, we see that as long as
$1-\psi>\frac{1}{\log n}$ we can choose $\eta=\frac{1}{(1-\psi)\log
n}$, yielding the probabilistic estimate
$$
\Pr\left[\forall\ k\in \mathbb N,\
B_X\left(x,\frac{1}{96\alpha}\cdot 8^{-k}\diam(X)\right)\subseteq
\mathscr R_k(x)\right]\ge  \Omega\left(\frac{1}{(1-\psi)\log n}
\cdot\frac{1}{n^{1-\psi}}\right).
$$
This estimate is better than Proposition~\ref{prop:ramtopart} when
$\frac{1}{\log n}<1-\psi<O\left(\frac{1}{\sqrt{\log n}}\right)$.}
\end{remark}

\bibliographystyle{abbrv}
\bibliography{ramseypart}

\def\cprime{$'$}
\begin{thebibliography}{10}

\bibitem{ALN05}
S.~Arora, J.~R. Lee, and A.~Naor.
\newblock Euclidean distortion and the sparsest cut.
\newblock In {\em STOC '05: Proceedings of the thirty-seventh annual ACM
  symposium on Theory of computing}, pages 553--562, New York, NY, USA, 2005.
  ACM Press.

\bibitem{AMNSW98}
S.~Arya, D.~M. Mount, N.~S. Netanyahu, R.~Silverman, and A.~Y. Wu.
\newblock An optimal algorithm for approximate nearest neighbor searching.
\newblock {\em Journal of the ACM}, 45:891--923, 1998.

\bibitem{ABCP99}
B.~Awerbuch, B.~Berger, L.~Cowen, and D.~Peleg.
\newblock Near-linear time construction of sparse neighborhood covers.
\newblock {\em SIAM J. Comput.}, 28(1):263--277, 1999.

\bibitem{Bartal96}
Y.~Bartal.
\newblock Probabilistic approximations of metric space and its algorithmic
  application.
\newblock In {\em 37th Annual Symposium on Foundations of Computer Science},
  pages 183--193, Oct. 1996.

\bibitem{BLMN05}
Y.~Bartal, N.~Linial, M.~Mendel, and A.~Naor.
\newblock On metric {R}amsey type phenomena.
\newblock {\em Ann. of Math. (2)}, 162(2):643--709, 2005.

\bibitem{BF00}
M.~A. Bender and M.~{Farach-Colton}.
\newblock The lca problem revisited.
\newblock In {\em Proc. 4th Latin Amer. Symp. on Theor. Info.}, pages 88--94.
  Springer-Verlag, 2000.

\bibitem{BF04b}
M.~A. Bender and M.~Farach-Colton.
\newblock The level ancestor problem simplified.
\newblock {\em Theoretical Comput. Sci.}, 321(1):5--12, 2004.

\bibitem{BFM86}
J.~Bourgain, T.~Figiel, and V.~Milman.
\newblock On {H}ilbertian subsets of finite metric spaces.
\newblock {\em Israel J. Math.}, 55(2):147--152, 1986.

\bibitem{CKR04}
G.~Calinescu, H.~Karloff, and Y.~Rabani.
\newblock Approximation algorithms for the 0-extension problem.
\newblock {\em SIAM J. Comput.}, 34(2):358--372 (electronic), 2004/05.

\bibitem{Call95}
P.~B. Callahan.
\newblock {\em Dealing with higher dimensions: the well-separated pair
  decomposition and its applications}.
\newblock Ph.{D}. thesis, Dept. Comput. Sci., Johns Hopkins University,
  Baltimore, Maryland, 1995.

\bibitem{CK95}
P.~B. Callahan and S.~Kosaraju.
\newblock {A decomposition of multidimensional point sets with applications to
  $k$-nearest-neighbors and $n$-body potential fields.}
\newblock {\em J. ACM}, 42(1):67--90, 1995.

\bibitem{CGR05}
S.~Chawla, A.~Gupta, and H.~R\"acke.
\newblock Embeddings of negative-type metrics and an improved approximation to
  generalized sparsest cut.
\newblock In {\em SODA '05: Proceedings of the sixteenth annual ACM-SIAM
  symposium on Discrete algorithms}, pages 102--111, Philadelphia, PA, USA,
  2005. Society for Industrial and Applied Mathematics.

\bibitem{Clarkson05}
K.~L. Clarkson.
\newblock Nearest-{N}eighbor {S}earching and {M}etric {S}pace {D}imensions.
\newblock In {\em Nearest-{N}eighbor {M}ethods for {L}earning and {V}ision:
  {T}heory and {P}ractice}. MIT Press.
\newblock Available at
  \url{http://cm.bell-labs.com/who/clarkson/nn_survey/p.pdf}, 2005.

\bibitem{Cohen99}
E.~Cohen.
\newblock Fast algorithms for constructing t-spanners and paths with stretch t.
\newblock {\em SIAM J. Comput.}, 28(1):210--236, 1999.

\bibitem{CLR90}
T.~H. Cormen, C.~E. Leiserson, and R.~L. Rivest.
\newblock {\em Introduction to Algorithms}.
\newblock MIT Press, 1990.

\bibitem{Erd64}
P.~Erd{\H{o}}s.
\newblock Extremal problems in graph theory.
\newblock In {\em Theory of Graphs and its Applications (Proc. Sympos.
  Smolenice, 1963)}, pages 29--36. Publ. House Czechoslovak Acad. Sci., Prague,
  1964.

\bibitem{FRT04}
J.~Fakcharoenphol, S.~Rao, and K.~Talwar.
\newblock A tight bound on approximating arbitrary metrics by tree metrics.
\newblock {\em J. Comput. System Sci.}, 69(3):485--497, 2004.

\bibitem{FK05}
C.~Fefferman and B.~Klartag.
\newblock Fitting ${C}^m$ smooth functions to data {I}.
\newblock Preprint, availabe at\\
  \url{http://www.math.princeton.edu/facultypapers/Fefferman/FittingData\_Part%
\_I.pdf}, 2005.

\bibitem{Gud02}
J.~Gudmundsson, C.~Levcopoulos, G.~Narasimhan, and M.~Smid.
\newblock Approximate distance oracles for geometric graphs.
\newblock In {\em SODA '02: Proceedings of the thirteenth annual ACM-SIAM
  symposium on Discrete algorithms}, pages 828--837, Philadelphia, PA, USA,
  2002. Society for Industrial and Applied Mathematics.

\bibitem{HM05}
S.~Har-Peled and M.~Mendel.
\newblock Fast construction of nets in low dimensional metrics, and their
  applications.
\newblock In {\em SCG '05: Proceedings of the twenty-first annual symposium on
  Computational geometry}, pages 150--158, New York, NY, USA, 2005. ACM Press.
\newblock Available at \url{http://arxiv.org/abs/cs.DS/0409057}, to appear in
  SIAM J. Computing.

\bibitem{HT84}
D.~Harel and R.~E. Tarjan.
\newblock Fast algorithms for finding nearest common ancestors.
\newblock {\em SIAM J. Comput.}, 13(2):338--355, 1984.

\bibitem{Hei01}
J.~Heinonen.
\newblock {\em Lectures on analysis on metric spaces}.
\newblock Universitext. Springer-Verlag, New York, 2001.

\bibitem{Ind04-handbook}
P.~Indyk.
\newblock Nearest neighbors in high-dimensional spaces.
\newblock In {\em Handbook of discrete and computational geometry, second
  edition}, pages 877--892. CRC Press, Inc., Boca Raton, FL, USA, 2004.

\bibitem{KLMN05}
R.~Krauthgamer, J.~R. Lee, M.~Mendel, and A.~Naor.
\newblock Measured descent: A new embedding method for finite metrics.
\newblock {\em Geom. Funct. Anal.}, 15(4):839--858, 2005.

\bibitem{LN05-via}
J.~R. Lee and A.~Naor.
\newblock {Extending Lipschitz functions via random metric partitions.}
\newblock {\em Invent. Math.}, 160(1):59--95, 2005.

\bibitem{Mat96}
J.~Matou\v{s}ek.
\newblock On the distortion required for embedding finite metric space into
  normed spaces.
\newblock {\em Israel J. Math.}, 93:333--344, 1996.

\bibitem{MN04}
M.~Mendel and A.~Naor.
\newblock Euclidean quotients of finite metric spaces.
\newblock {\em Adv. Math.}, 189(2):451--494, 2004.

\bibitem{Orc91}
M.~T. Orchard.
\newblock A fast nearest-neighbor search algorithm.
\newblock In {\em ICASSP '91: Int. Conf. on Acoustics, Speech and Signal
  Processing}, volume~4, pages 2297--3000, 1991.

\bibitem{Peleg04}
D.~Peleg.
\newblock Informative labeling schemes for graphs.
\newblock {\em Theoretical Computer Science}, 340(3):577--593, 2005.

\bibitem{LTZ05}
L.~Roditty, M.~Thorup, and U.~Zwick.
\newblock Deterministic constructions of approximate distance oracles and
  spanners.
\newblock In {\em Automata, languages and programming}, volume 3580 of {\em
  Lecture Notes in Comput. Sci.}, pages 261--272. Springer, Berlin, 2005.

\bibitem{Tal04}
K.~Talwar.
\newblock Bypassing the embedding: algorithms for low dimensional metrics.
\newblock In {\em STOC '04: Proceedings of the thirty-sixth annual ACM
  symposium on Theory of computing}, pages 281--290, New York, NY, USA, 2004.
  ACM Press.

\bibitem{TZ05}
M.~Thorup and U.~Zwick.
\newblock Approximate distance oracles.
\newblock {\em J. ACM}, 52(1):1--24, 2005.

\bibitem{Vai89}
P.~M. Vaidya.
\newblock An {$O(n\log n)$} algorithm for the all-nearest-neighbors problem.
\newblock {\em Discrete Comput. Geom.}, 4(2):101--115, 1989.

\bibitem{Z01}
U.~Zwick.
\newblock Exact and approximate distances in graphs---a survey.
\newblock In {\em Algorithms---ESA 2001 (\AA rhus)}, volume 2161 of {\em
  Lecture Notes in Comput. Sci.}, pages 33--48. Springer, Berlin, 2001.

\end{thebibliography}

\bigskip
\bigskip
\bigskip

\noindent Manor Mendel. Computer Science Division, The Open
University of Israel, 108 Ravutski Street P.O.B. 808, Raanana
43107, Israel. Email: {\tt mendelma@gmail.com}.

\bigskip

\bigskip

\noindent Assaf Naor. One Microsoft Way, Redmond WA, 98052, USA.
Email: {\tt anaor@microsoft.com}.

\end{document}